\documentclass[12pt,a4paper]{article}

\usepackage{amsmath}
\usepackage{amssymb}
\usepackage{graphicx}
\usepackage{float}

\usepackage[dvips]{color}

\usepackage{colordvi}

\usepackage{url}

\begin{document}

\def\gap#1{\vspace{#1 ex}}
\def\be{\begin{equation}}
\def\ee{\end{equation}}
\def\bal{\begin{array}{l}}
\def\ba#1{\begin{array}{#1}}  
\def\ea{\end{array}}
\def\bea{\begin{eqnarray}}
\def\eea{\end{eqnarray}}
\def\beas{\begin{eqnarray*}}
\def\eeas{\end{eqnarray*}}
\def\del{\partial}
\def\eq#1{(\ref{#1})}
\def\fig#1{Fig \ref{#1}} 
\def\re#1{{\bf #1}}
\def\bull{$\bullet$}
\def\nn{\\\nonumber}
\def\ub{\underbar}
\def\nl{\hfill\break}
\def\ni{\noindent}
\def\bibi{\bibitem}
\def\ket{\rangle}
\def\bra{\langle}
\def\vev#1{\langle #1 \rangle} 
\def\lsim{\stackrel{<}{\sim}}
\def\gsim{\stackrel{>}{\sim}}
\def\mattwo#1#2#3#4{\left(
\begin{array}{cc}#1&#2\\#3&#4\end{array}\right)} 
\def\tgen#1{T^{#1}}
\def\half{\frac12}
\def\floor#1{{\lfloor #1 \rfloor}}
\def\ceil#1{{\lceil #1 \rceil}}

\def\mysec#1{\gap1\ni{\bf #1}\gap1}
\def\mycap#1{\begin{quote}{\footnotesize #1}\end{quote}}

\def\bit{\begin{item}}
\def\eit{\end{item}}
\def\benu{\begin{enumerate}}
\def\eenu{\end{enumerate}}
\def\a{\alpha}
\def\as{\asymp}
\def\ap{\approx}
\def\b{\beta}
\def\bp{\bar{\partial}}
\def\cA{{\cal{A}}}
\def\cD{{\cal{D}}}
\def\cL{{\cal{L}}}
\def\cP{{\cal{P}}}
\def\cR{{\cal{R}}}
\def\da{\dagger}
\def\de{\delta}
\def\tD{\tilde D}
\def\e{\eta}
\def\ep{\epsilon}
\def\eqv{\equiv}
\def\f{\frac}
\def\g{\gamma}
\def\G{\Gamma}
\def\h{\hat}
\def\hs{\hspace}
\def\i{\iota}
\def\k{\kappa}
\def\lf{\left}
\def\l{\lambda}
\def\la{\leftarrow}
\def\La{\Leftarrow}
\def\Lla{\Longleftarrow}
\def\Lra{\Longrightarrow}
\def\L{\Lambda}
\def\m{\mu}
\def\na{\nabla}
\def\nn{\nonumber\\}
\def\mm{&&\kern-18pt}  
\def\om{\omega}
\def\O{\Omega}
\def\P{\Phi}
\def\pa{\partial}
\def\pr{\prime}
\def\r{\rho}
\def\ra{\rightarrow}
\def\Ra{\Rightarrow}
\def\ri{\right}
\def\s{\sigma}
\def\sq{\sqrt}
\def\S{\Sigma}
\def\si{\simeq}
\def\st{\star}
\def\t{\theta}
\def\ta{\tau}
\def\ti{\tilde}
\def\tm{\times}
\def\tr{\textrm}
\def\Tr{{\rm Tr}}
\def\T{\Theta}
\def\up{\upsilon}
\def\Up{\Upsilon}
\def\v{\varepsilon}
\def\vh{\varpi}
\def\vk{\vec{k}}
\def\vp{\varphi}
\def\vr{\varrho}
\def\vs{\varsigma}
\def\vt{\vartheta}
\def\w{\wedge}
\def\z{\zeta}

\thispagestyle{empty}
\addtocounter{page}{-1}
{}
\vskip-5cm
\begin{flushright}
TIFR/TH/11-10\\
CCTP-2011-07\\
\end{flushright}
\vspace*{0.1cm} \centerline{\Large \bf 
Phases of a two dimensional large-$N$ gauge theory on a torus  }
\vspace*{1 cm} 
\centerline{\bf 
Gautam~Mandal$^1$ and Takeshi~Morita$^2$}
\vspace*{0.5cm}
\centerline{\rm $^1$\it Department of Theoretical Physics,}
\centerline{\it Tata Institute of Fundamental Research,} 
\centerline{\it Mumbai 400 005, \rm INDIA}
\vspace*{0.3cm}
\centerline{\rm $^2$\it Crete Center for Theoretical Physics}
\centerline{\rm \it Department of Physics}
\centerline{\it University of Crete, 71003 Heraklion, \rm Greece}
\vspace*{0.3cm}
\centerline{\tt email: mandal@theory.tifr.res.in, takeshi@physics.uoc.gr
}

\vspace*{0.4cm}
\centerline{\bf Abstract}
\vspace*{0.3cm} 

\enlargethispage{1000pt}

We consider two-dimensional large $N$ gauge theory with $D$ adjoint
scalars on a torus, which is obtained from a $D+2$ dimensional pure
Yang-Mills theory on $T^{D+2}$ with $D$ small radii. The two
dimensional model has various phases characterized by the holonomy of
the gauge field around non-contractible cycles of the 2-torus.  
We determine the phase boundaries and derive the order of the phase transitions using a method, developed in an earlier work (hep-th/0910.4526), which is nonperturbative in the 'tHooft coupling and uses a $1/D$ expansion.  
We embed our phase diagram in the more
extensive phase structure of the $D+2$ dimensional Yang-Mills theory
and match with the picture of a cascade of phase transitions found
earlier in lattice calculations (hep-lat/0710.0098). We also propose a
dual gravity system based on a Scherk-Schwarz compactification of a D2
brane wrapped on a 3-torus and find a phase structure which is similar
to the phase diagram found in the gauge theory calculation.

\newpage

\tableofcontents

\section{Introduction and Summary}

Gauge theories on spaces with compact directions have been studied for
a long time. As a prototypical example, $d+1$ dimensional Yang-Mills
theory at a finite temperature $T$ corresponds to a compactification
of the (Euclidean) time direction on a circle of length
$\beta=1/T$. It is obviously important to study such a
compactification to understand the physics of
confinement/deconfinement transitions \cite{Gross:1980br}. More
generally, one can consider Yang-Mills theory on a compact space
$\Sigma$.  If the volume of $\Sigma$ is finite, there is no sharp
phase transition, but for an $SU(N)$ gauge theory in the large $N$
limit there are sharply demarcated phases depending on the shape and
size parameters of $\Sigma$.  In case the compact space is a torus,
the phase diagram as a function of various radii (and coupling)
reveals a rich phase structure \cite{Narayanan:2003fc,
  Aharony:2005ew}, including a cascade of phase transitions in which
the ``Polyakov'' loops along various non-contractible cycles become
non-zero in succession as the radii are reduced
\cite{Narayanan:2007fb, Hanada:2007wn}. Most of these studies are
numerical (in the lattice or in the continuum) or, in some cases,
based on holography (see Section \ref{other} for references and more
details). One of the motivations of the present paper is to
investigate these questions analytically in a simple situation, as
explained below, by using and extending the ``large $D$'' technique
developed in \cite{ Mandal:2009vz}.

To elaborate further, let us consider a Euclidean $d+D$-dimensional
gauge theory \footnote{In this paper, we will not consider the
  contribution of the $\theta$ term.}  on a $d+D$-dimensional torus with
radii $L_\mu$ \footnote{Our notation for spacetime coordinates is: $\{
  x^0 \equiv t, x^M\}, M=1,..., d+D-1$. We will further split the
  $d+D-1$ coordinates into $d$ `large' dimensions $\{ x_0, x^i \},
  i=1,..., d-1$ and $D$ `small' dimensions $x^I, I=1,2,...D$ (the
  meaning of `large' and `small' is explained below).}
\begin{align} 
S= \int_{0}^\beta \kern-5pt dt \left(  \prod_{M=1}^{D+d-1}
\int_0^{L_M}\kern-5pt dx^M \right) \frac{1}{4g^2_{d+D}}\Tr F_{\mu\nu}^2 
\label{d=d+D}.
\end{align} 
Here the length of the temporal circle is denoted as $L_0=\b$ and the
rest are denoted as $L_M, M=1,..., d+D-1$. The phases of \eq{d=d+D}
are characterized by Wilson lines around the $d+D$ noncontractible
cycles of the torus:
\begin{align}
W_\mu = \Tr U_\mu  \equiv 
\frac{1}{N}\Tr P \left(\exp \left[ i \int_0^{L_\mu} A_\mu dx^\mu \right]
\right),
\end{align}
where no sum over $\mu$ is intended. These Wilson loops transform
nontrivially under the centre symmetry\footnote{ \label{ftnt-centre}
  `Centre symmetry'\cite{Luscher:1982ma,'tHooft:1979uj} is generated
  by quasiperiodic `gauge transformations' $\alpha(x^\mu) = $ $
  \exp[2\pi i (n_\mu x^\mu/L_\mu) {\cal A}]$, where ${\cal A}= {\rm
    diag}[1/N,1/N,...,1/N,(1-N)/N]$. The quasiperiodicity is up to
  phases $h_\mu = \exp[2\pi i n_\mu /N]$, $\mu=0,1,...,d+D-1$ which
  parametrize $d+D$ copies of the centre of $SU(N)$, $Z_N^{d+D}$. The
  $\a(x^\mu)$ are valid gauge transformations locally, and leave local
  colour-singlets e.g.  tr$F_{\mu\nu}^2$ invariant; in particular they
  commute with the hamiltonian. However, under the
  $\alpha$-transformations $W_\mu \to h_\mu W_\mu$. A non-zero value
  of $\vev{W_\mu}$ implies spontaneous symmetry breaking of the centre
  symmetry in the $\mu$-direction.}. For sufficiently large radii
$L_\mu$, all $W_\mu$ vanish, signifying unbroken centre symmetry. 
In this phase, local gauge-invariant observables are independent of $L_\mu$ in the strict large $N$ limit \cite{Eguchi:1982nm, Gocksch:1982en, Unsal:2010qh}. 
Since $W_0$ can be interpreted as $\exp[- S_q]$, where $S_q$ is the action for a static quark, the phase with
$\vev{W_0}=0$ exhibits confinement. As is well-known, as $\b$ is
reduced (i.e. the temperature is increased), below a certain critical
value $\b_c$, $\vev{W_0}$ becomes non-zero, signaling a deconfinement
transition together with a breaking of the centre symmetry $Z_N^{d+D}
\to Z_N^{d+D-1}$. 
In this phase, the observables can depend on $\beta$ but are still 
independent of $L_i$ \cite{Gocksch:1982en}.
 It has been argued from lattice studies (see
\cite{Narayanan:2007fb} and Section \ref{other} for a review) that as
the other radii are successively reduced, one has a cascade of
analogous symmetry breaking transitions $Z_N^{d+D-1} \to Z_N^{d+D-2}
\to ... \to 1$.

While it would be fascinating to study all the above phases
analytically, in this paper we will be able to study the phases of a
$D+2$ dimensional pure Yang Mills theory on $T^{D+2}$ (i.e. \eq{d=d+D}
with $d=2$) in which a $D$-dimensional torus (with radii $L^I/(2\pi),
I=1, 2, ..., D$) is taken as small (ensuring broken $Z_N$ symmetries
in those directions), leaving the remaining $d=2$ directions
(including time) of variable size. Such a theory is given by a
Kaluza-Klein reduction \footnote{\label{tricky} Kaluza Klein reduction
  is tricky for gauge theories \cite{Aharony:2005ew, Unsal:2010qh}, since in the
  confined phase the KK modes can have energies $\sim 1/(NL)$, which
  become arbitrarily low at large $N$. The fractional modes,
  equivalent to the `long string' modes of \cite{Dijkgraaf:1997vv},
  can be understood as arising from mode shifts of charged fields in
  the presence of Wilson lines whose eigenvalues are uniformly
  distributed along a circle (see Section \ref{sec-large-L} for an
  explicit verification for this statement). In the deconfined phase,
  however, the KK modes have energies $\sim 1/L$, like in ordinary
  field theories, and KK reduction proceeds as usual.}  of \eq{d=d+D}
on the small $T^D$, and is described by the following action:
\begin{align} 
\kern-5pt S = \int_{0}^\beta \kern-5pt dt \int_0^{L}\kern-5pt dx \,
\Tr \Biggl( & \frac{1}{2{g}^2} F_{01}^2 + \sum_{I=1}^D \frac12
\left(D_\mu Y^{I}\right)^2+\frac{m^2}{2}(Y^I)^2 \nn & - \sum_{I,J}
\frac{{g'}^2}{4} [Y^I,Y^J][Y^I, Y^J] \Biggr).
\label{d=2}
\end{align}
Here $Y^I$ comes from the gauge field components $A_{I+1}$ and the
covariant derivative is defined as $D_\mu = \del_\mu - i[A_\mu,
  ~~]$. A naive KK reduction leads to massless $Y^I$'s and $g=g'$;
however, a mass $m$ for the adjoint scalars as well as radiative
splitting between $g$ and $g'$ is induced from loops of KK modes (see
Appendix \ref{KK loop} and Section \ref{sec-large-L}, respectively,
for more details).

We should remark that \eq{d=2} can either be regarded as a step
towards understanding the full phase diagram of the $d+D$ dimensional
theory \eq{d=d+D}, or be understood as a two-dimensional gauge
theory in its own right. The spirit of the latter approach is to
provide an example of an analytically solvable low-dimensional gauge
theory in the limit of a large number of adjoint scalars (the $d=1$
theory was discussed in \cite{Mandal:2009vz}). Equation \eq{d=2}, from
this viewpoint, provides a bosonic counterpart of the Gross-Neveu
model \cite{Gross:1974jv} where $D$ plays the role of $N_f$, and the
$SO(D)$-invariant bilinear $\sum_{I=1}^D Y^I_a Y^I_b$ of Section
\ref{sec-large-D} plays the role of $SU(N_f)$-invariant fermion
bilinears such as $\sum_{i=1}^{N_f} \bar \psi_i \psi_i$ of the
Gross-Neveu model.

Note that in this second point of view, where we regard the action
(\ref{d=2}) as an independent theory in its own right, the mass $m$
can be taken to be arbitrary. In particular, if the mass is
sufficiently large ($m^2 \gg g^2ND, g'^2ND$), the phase structure can
be determined perturbatively \cite{Semenoff:1996xg, Aharony:2005ew}.
However, as we will see in appendix \ref{KK loop}, if we regard
\eq{d=2} as a KK reduction of \eq{d=d+D} on $T^D$, then the mass $m$ of
the adjoint scalars is much lower than the scales mentioned above and
the theory is not amenable to such perturbative methods. One of the
goals of the present work is to provide a nonperturbative \footnote{in
  the 'tHooft coupling.} analysis of \eq{d=2} valid for any value of
mass (including $m=0$), based on the `large $D$' method developed in
the previous work \cite{Mandal:2009vz}.

A few additional comments are in order:

(i) KK gauge theories have important applications to phenomenology
\cite{ArkaniHamed:1998rs,ArkaniHamed:1998nn, Randall:1999ee,
  Randall:1999vf}. Theories such as \eq{d=2} provide important toy
models in this context. In particular, issues such as different
running of gauge couplings in the compactified and decompactified
theories can be examined in such models. We will encounter some of
these issues in Section \ref{sec-large-L}.

(ii) Gauge theories on compact spaces can sometimes have gravity duals.
The deconfinement transition in ${\cal N}=4$ super Yang Mills theory
on $S^3 \times S^1$ \cite{Sundborg:1999ue, Aharony:2003sx} is a weak
coupling continuation of the gravitational Hawking-Page transition
\cite{Hawking:1982dh, Witten:1998zw}. Similar correspondences for
two-dimensional supersymmetric gauge theories on tori were analyzed in
\cite{Aharony:2004ig, Aharony:2005ew}\footnote{For an extensive list
  of correspondences between low dimensional gauge theories and
  gravitational systems, see \cite{Itzhaki:1998dd, Martinec:1998ja,
    Aharony:1999ti}.}. In this paper, we will obtain \eq{d=2} with
$D=8$ from a Scherk-Schwarz compactification of a three dimensional
super Yang Mills theory with sixteen supercharges, which corresponds
to the world volume theory of D2 branes. The latter theory has an
AdS/CFT dual \cite{Itzhaki:1998dd}, which leads to the construction of
a gravity dual for \eq{d=2} in a sense defined in Section
\ref{sec-gravity} \footnote{One of the motivations for this work was to
  construct a gauge theory dual to a dynamical Gregory-Laflamme
  transition.  We will discuss this issue in a forthcoming publication
  \cite{progress}.}. As we will see, the gravity analysis will complement
our knowledge of the phase structure from the gauge theory analysis.

(iii) Large $N$ two dimensional gauge theories themselves are
interesting objects in the context of string theory and QCD.  For
instance, confinement/deconfinement type transitions
\cite{Polchinski:1991tw} have been analytically found in 2D
QCD with heavy adjoint scalars \cite{Semenoff:1996xg}. In addition, stringy
excitations and glueball spectra have been obtained in 2D models in
\cite{Dalley:1992yy, Kutasov:1993gq, Bhanot:1993xp, Demeterfi:1993rs,
  Dhar:1994aw}.  Thus, these models are good laboratories for real QCD.
Our study, in fact, has a direct relevance for \cite{Demeterfi:1993rs};
we hope to return to the issue of glueball spectrum discussed in this
reference.

The principal result in this paper is the determination of some parts
of the phase diagram of the two-dimensional theory (\ref{d=2}) at weak
coupling.  The result is summarized in figure \ref{fig-phases}.  The
second result is the gravity analysis which complements the phase
diagram at strong coupling, which is presented in figure
\ref{fig-phases-gr}.

The plan of the paper is as follows.

In Sections \ref{sec-large-L}, \ref{sec-small-L} and
\ref{sec-phase-gauge} we analyse the model \eq{d=2} at weak coupling
by using the $1/D$ expansion \cite{Hotta:1998en} developed in
\cite{Mandal:2009vz}.  We find that the nature and the order of the
confinement/deconfinement type transition depends on whether the size
$L$ of the spatial circle is large or small. For large $L$ (which
corresponds to the $\Tr V=0$ phase), we find (see Section
\ref{sec-large-L}) a single first order transition, thus providing
analytical evidence for earlier lattice studies (see Section
\ref{lattice} for further details). On the other hand for small $L$
(corresponding to the $\Tr V\not= 0$ phase), the analysis in
\cite{Mandal:2009vz} is valid and, as detailed in Section
\ref{sec-small-L}, the transition consists of two higher order phase
transitions. The phase diagram is summarised in Section
\ref{sec-phase-gauge} in figure \ref{fig-phases} and is in agreement
with those from the lattice studies of \cite{Narayanan:2003fc,
  Hanada:2007wn, Narayanan:2007fb}. We compare our results with these
lattice studies and with \cite{Aharony:2005ew} in Section \ref{other}.

To supplement our $1/D$ analysis of the gauge theory, we consider in
Section \ref{sec-gravity} a dual gravity theory, obtained from  D2
branes wrapped on a 3-torus with a Scherk Schwarz circle.  Although the
gravity results pertain to thermodynamics in the strongly coupled
regime, we observe that the phase structure is qualitatively similar
to that of the weak coupling gauge theory. This allows us to arrive at
a conjectured phase diagram in figure \ref{fig-phases-gr}, which
suggests a particular way of connecting the phase boundaries of
figure \ref{fig-phases}.

In Section \ref{topology} we comment on the dependence of the order of
phase transition on the topology of the compact space.  In Appendix
\ref{KK loop} we discuss the masses of the adjoint scalars which
appear from integration of the KK modes at the one loop level.  In
Appendix \ref{app-derivation} we fill in some details needed in
Section \ref{sec-large-L} for integrating out the adjoint scalars.  In
Appendix \ref{app-massive} we discuss the influence of the mass term
in (\ref{d=2}) on the phase diagram.  In Appendix \ref{app-grav} we
provide important details of our gravity analysis.

\section{Confinement/deconfinement type transition in large radius torus}
\label{sec-large-L}

In this section, we will analyse the confinement/deconfinement type
transition in (\ref{d=2}) for large $L$. First, (in Section
\ref{sec-large-D}) we will integrate out the adjoint scalars $Y^I$ in
a $1/D$ expansion, in a manner similar to \cite{Mandal:2009vz},
leading to the effective hamiltonian \eq{eff-action} in the large $L$
limit. Next (in Section \ref{sec-phase tr}) we use , at $L \to
\infty$, the results of
\cite{Semenoff:1996xg, Gattringer:1996fi, Basu:2006mq}, who studied
this effective action in slightly different contexts, to determine the phase structure of our theory \eq{d=2} at
large $L$. The justification for extrapolating their result to finite
$L$, as detailed below, comes from the phenomenon of large $N$ volume independence
\cite{Eguchi:1982nm, Gocksch:1982en, Unsal:2010qh} which is valid as
long as $L$ is large enough to ensure $\Tr V=0$.

To keep the analysis simple, in this section we consider \eq{d=2} with
$m=0$, and defer the case of nonzero mass to appendix
\ref{app-massive}. As we will find there, the inclusion of the mass
term does not change the qualitative structure of the different
phases.

\subsection{Large $D$ saddle point}
\label{sec-large-D}

In this section, we will generalize the analysis of the
$0+1$-dimensional gauge theory \cite{Mandal:2009vz} to the $d=2$ model \eq{d=2} and show
that if we consider the number $D$ of adjoint scalars to be large, the
theory can be considered to be in the vicinity of a large $D$ saddle
point \footnote{Although the saddle point arises in a manner similar
  to that in four-fermion models such as Gross-Neveu or Nambu
  Jona-Lasinio, the saddle point is {\em complex}. See
  \cite{Mandal:2009vz} for details.} and various quantities such as
the free energy and the mass gap etc. can be computed around the
saddle point in a $1/D$ expansion.

As mentioned above, in this section we will consider \eq{d=2} with
$m=0$.  Introducing an auxiliary field $B_{ab}$, the path-integral of
the gauge theory can be rewritten as
\begin{align} 
Z= & {\cal N} \int {\cal D}B {\cal D}A_\mu {\cal D}Y^I e^{-S(B,A,Y)},
\nn 
\kern-7pt S(B,A,Y) &=\int_0^\beta \kern-7pt dt \int_0^L \kern-7pt dx
\left[ \frac{1}{2{g}^2}  
{F_{01}^a}^2+ \frac12 \left(D_\mu Y^{I}_a\right)^2 - i\frac12 B_{ab}
  Y_{a}^I Y_{b}^I + \frac{1}{4{g'}^2} B_{ab} M^{-1}_{ab,cd}B_{cd}
  \right],
\label{gauss-trick}
\end{align} 
where ${\cal N}$ is a constant factor and we have used the following
matrix,
\begin{align} 
M_{ab,cd} = -\frac{1}{4} \Bigl\{ \Tr[\l_a,
  \l_c][\l_b, \l_d] +(a\leftrightarrow b)+(c\leftrightarrow
d)+(a\leftrightarrow b,c\leftrightarrow d) \Bigr\},
\label{def Mabcd}
\end{align}
$\lambda^a$ ($a=1,\cdots,N^2-1$) being the generators of $SU(N)$.  Our
approach, similar to the one-dimensional case, will be as follows. We
will integrate out the $Y^I$'s to obtain an effective action for
$A_\mu$ and $B_{ab}$, and find a saddle point solution for $B_{ab}$ (for given
$A_\mu$) in a large $D$ limit. The effective action for $A_\mu$ will
be essentially obtained by substituting the saddle point value of $B_{ab}$
in this effective action.

As in \cite{Mandal:2009vz}, it is convenient to decompose $B_{ab}$ as
the sum of a trace piece (independent of $x,t$) and an orthogonal
part:
\begin{align}
B_{ab}(t) = i \Delta^2 \delta_{ab} +g' b_{ab}(t,x),
\label{fluct}
\end{align}
where $b_{ab}$ satisfies $\int dt \int dx~ b_{aa}=0$.  Such a
decomposition, into a large diagonal piece and a small off-diagonal
fluctuation, will be {\em a posteriori} justified by finding a saddle
point for $B_{ab}$ of the form $\langle B_{ab} \rangle = i
{\Delta_0}^2 \delta_{ab}$, where $\Delta_0$ is a real constant
(depending only on the 'tHooft coupling).

With this decomposition, the action reduces to
\begin{align}
S_{0}&= -\frac{\beta L N \Delta^4}{8{g'}^2} , \nonumber \\ S_1
&=\int_0^\beta dt \int_0^L dx \left( \frac{1}{2{g}^2}
{F_{01}^a}^2+\frac{1}{4} b_{ab} M_{ab,cd}^{-1} b_{cd} + \frac12
\left(D_\mu Y_a^{I}\right) ^2 + \frac{1}{2} \Delta^2
Y^{I2}_{a}\right) ,\nonumber \\ S_{int}&=- \int_0^\beta dt \int_0^L dx
\left(\frac{ig'}{2} b_{ab}Y^I_aY^I_b \right) ,
\label{action 1dMM}
\end{align} 
where we have used $M_{ab,cd}^{-1}\delta_{cd}=\delta_{ab}/2N$ \cite{Mandal:2009vz}.
Let us now take a large $D$ limit,
\begin{align} 
g,g' \to 0, \quad N,D \to \infty \quad s.t. \quad
\tilde{\lambda}\equiv g^2DN, \quad \tilde{\lambda'}\equiv {g'}^2DN
\quad \text{fixed}.
\label{thooft-limit}
\end{align} 
To the leading order of this expansion, we can ignore the interaction
term $S_{int}$.\footnote{In the one dimensional model ($d=1$), the
  next order of the $1/D$ expansion has been evaluated in
  \cite{Mandal:2009vz}.  There, such $1/D$ corrections do not change
  the nature of the phase structure.  We can expect that the same
  thing will happen in our two dimensional gauge theory also.  Hence
  we do not evaluate the $1/D$ corrections in this article.}  In that
case, the integration of $b_{ab}$ will contribute just a numerical
factor and we will ignore it. Integrating over the $Y^I$'s in $S_1$,
we then get the following leading result for the partition function
\begin{align}
Z &=  \int {\cal D}A {\cal D}\Delta\  e^{- S_{eff}[A, \Delta]},
\nn
S_{eff}[A, \Delta] &= -\frac{\beta L N \Delta^4}{8{g'}^2} +
\int_0^\beta dt \int_0^L dx  \frac{1}{2{g}^2} {F_{01}^a}^2 +
\delta S_{eff},
\label{s-eff}
\end{align}
where $\delta S_{eff}$ is the 1-loop contribution from the
$Y^I$-integration:
\begin{align}
\delta S_{eff}[A, \Delta]=\frac{D}{2} \log \det \left( -D^2_\mu
+\Delta^2 \right).
\label{logdet}
\end{align} 
It is difficult to evaluate this last quantity in general.  However,
using arguments similar to \cite{Aharony:2005ew}, we will find that:

\gap1

\noindent (a) If $\Delta^2 \gg \tilde{\lambda} \equiv {g}^2 DN$ (this
assumption will be justified in \eq{choice2}), terms in $\delta
S_{eff}$ involving derivatives of $A_\mu$, e.g. terms involving the
gauge field strength and its covariant derivatives are suppressed.
\hfill\break 
(b) If $L \Delta \gg 1$,\footnote{This assumption will be justified later in what
we will define as the `large $L$ regime' $L \gtrsim
L_c$ (see \eq{large-L} and \eq{beta-2}), since $L_c \Delta $ will
turn out be large, using \eq{choice2}.}
we can approximately treat $A_0$ and $A_1$ as commuting matrices.

The argument for assertion (a) can be sketched briefly as
follows. Consider the simplest of the 1-loop diagrams
(figure \ref{fig-feynman}) which contribute to \eq{logdet} and has only
two external gauge field insertions:
\begin{figure}[H]
\begin{center}
\includegraphics[scale=0.6]{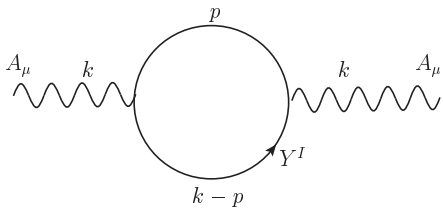}
\caption{A simple Feynman diagram contributing to \eq{logdet}.}
\label{fig-feynman}
\end{center}
\end{figure}
For large $\Delta$, the $Y^I$-propagator in the loop carrying
momentum $p$ can be expanded in powers of $p^2/\Delta^2$. The first
term in that expansion goes as $ND (k_\mu k_\nu - k^2
g_{\mu\nu})/\Delta^2$ (in the Feynman gauge), where the factors of
$N$ and $D$ come from the $Y^I$-loop, and $k$ is the external momentum
going into the loop. This term amounts to a correction, to the
$F_{01}^2/{g}^2$ term, of the form $(1+O(\tilde\l/\Delta^2))$, as
claimed above. We will ensure below that $\tilde\l/\Delta^2 \ll 1$.

The argument for assertion (b) will follow {\em a posteriori}
after we proceed with the assumption that $A_\mu$ are constant
commuting matrices. Under this assumption, as detailed
in appendix \ref{app-derivation}, the one-loop term becomes
\begin{align}
\delta S_{eff} = \frac{D \beta L}{8\pi^2} &
\Biggl[N^2
\left(
-\pi\Lambda^2 +\pi \left(  \Lambda^2+\Delta^2 \right)  \log
\left( \Lambda^2+\Delta^2 \right) 
-\pi \Delta^2 \log \Delta^2
\right) 
\nn
&-  \sum_{(k,l)\ne (0,0)}|\Tr(U^k V^l)|^2 
 \frac{4\pi
    \Delta}{\sqrt{(\beta k)^2+(Ll)^2}}K_1\left(\Delta
  \sqrt{(\beta k)^2+(Ll)^2} \right) \Biggr].
\label{arbit-u-v}
\end{align} 
Here we have used a momentum cut-off $\Lambda$ to regulate the
determinant\footnote{The cut off $\Lambda$ should be smaller than
  $M_{KK}$, which is the inverse length scale of the $D$-dimensional
  compactification torus.}, and used the notation $U=e^{i\beta A_0} $
and $V=e^{iLA_1}$. $K_1$ is the modified Bessel function of the second
kind. Using the asymptotic expansion $K_1(z)=\sqrt{\pi/2z}
e^{-z}+\cdots$ for large $z$ (justified below) and omitting some
irrelevant divergent terms, we get
\begin{align}
\delta S_{eff} = &\frac{D N^2 \beta L}{8\pi}
\left[\left(\Lambda^2+  \Delta^2 \right)\log\left(
1+\frac{\Lambda^2}{\Delta^2}  \right)  +\Delta^2\log  
\left(\frac{\Lambda^2}{\Delta^2} \right) \right] \nn &-
  \frac{D}{\sqrt{2\pi}} \left( L \sqrt{\frac{\Delta}{\beta} }
  e^{-\Delta \beta } \left| \Tr U \right|^2 + \beta
  \sqrt{\frac{\Delta}{L} } e^{-\Delta L } \left| \Tr V \right|^2
  \right) +\cdots,
\label{only-u}
\end{align}
where the $\cdots$ terms represent higher order terms in $e^{-\Delta
  \beta }, e^{-\Delta L }$. Since we are interested in large $L $, it
is obvious why higher order terms in $e^{-\Delta L }$ should be
ignored.  We now have an {\em a posteriori} justification for having
ignored terms involving commutators $[U,V]$; the smallest gauge
invariant such term would have at least 2 $U$s and 2 $V$s and hence
are expected to be of the same order as the $U^2V^2$ term in
\eq{arbit-u-v}, and hence can be ignored \cite{Aharony:2005ew}.
Higher order terms in $e^{-\Delta \beta }$ are ignored because they
will be small in the region of parameter space in which interesting
phase structures will appear.  We will ensure this at the end of
Section \ref{sec-phase tr} (see comment (a) below \eq{beta-2}).

In the $L \to \infty$ limit we can ignore the term in \eq{only-u}
involving $V$. Let us integrate $\Delta$ from \eq{s-eff}
under this assumption and derive the effective action for the
$U$-variable. The saddle point equation with respect to $\Delta^2$
reads as
\begin{align} 
-\frac{\beta L\Delta^2}{2\tilde{\lambda'} }
+\frac{\beta L }{4\pi}\log\left(1+\frac{\Lambda^2}{\Delta^2}   \right)  
+\frac{L}{\sqrt{2\pi}}\sqrt{\frac \beta \Delta}e^{-\Delta \beta}
\left(1-\frac{1}{2\Delta \beta}  \right) 
\left| \frac{1}{N} \Tr U \right|^2+\cdots=0.     
\label{saddle-point}
\end{align} 
In the $\Tr U=0$ phase (which is realized for sufficiently large
$\b$), $\Delta = \Delta_0$ is determined implicitly from the equation
\begin{align} 
\tilde{\lambda}' =\frac{2\pi \Delta_0^2}{\log\left(1+
  \frac{\Lambda^2}{\Delta_0^2} \right) }.
\label{delta-large-b}
\end{align} 
This equation can be viewed as a renormalization condition which
assigns a $\L$-dependence to $\tilde \l'$ such that the physical mass
scale $\Delta_0$ is held fixed. Here $\Delta_0$ plays a role analogous
to $\Lambda_{QCD}$ and its choice specifies the theory. We will, in
fact, choose it in such a way as to ensure the
condition \footnote{\label{ren-cond} This follows from the fact that
  the distinction between $g$ and $g'$ in \eq{d=2} vanishes in the
  original pure YM theory \eq{d=d+D}.  In order to fix the precise
  coefficient of the renormalization condition (\ref{choice}), we need
  to evaluate the contribution of the mass (\ref{saddle-massive}) and
  the running of $g$ and $g'$ for scales larger than $M_{KK}$.  The
  value of mass and the running of the couplings depend on the details
  of the higher dimensional theory. If we are interested in the two
  dimensional gauge theory (\ref{d=2}) itself as in the comment (iii)
  in the introduction, the renormalization condition that replaces
  (\ref{choice}) is arbitrary. However, even in that case, the
  qualitative nature of the phase structures in this paper does not
  change as long as $(\Delta_0^2+m^2)/ \tilde{\lambda} \gg 1$ is
  satisfied.}
\begin{align}
\tilde{\lambda}\sim \tilde{\lambda}'  \quad \text{at} ~ \Lambda=M_{KK}.
\label{choice}
\end{align}

As $\beta$ decreases, $\Tr U$ eventually becomes non-zero. However,
near criticality $\beta \Delta \gg 1$ (this is justified because of
\eq{choice2} and \eq{beta-2}).  Therefore, we can solve
\eq{saddle-point} for $\Delta$ in the form $\Delta_0 + O(\exp[-\b
  \Delta_0])$. We will write the explicit solution only for
$\Lambda\gg \Delta$:
\begin{align} 
\Delta=\Delta_0+\frac{1}{2\pi 
\Delta_0^2/\tilde{\lambda'} +1 }\sqrt{\frac{2\pi \Delta_0}{\beta} }
e^{-\Delta_0 \beta}  \left| \frac{1}{N} \Tr U \right|^2+\cdots
\label{delta-value},
\end{align} 
where
\begin{align} 
\Delta_0=
\sqrt{\frac{\tilde{\lambda'}}{2\pi} 
W\left(\frac{2\pi \Lambda^2}{
\tilde{\lambda'} }  \right)  }
\label{condensate},
\end{align} 
and $W(z)$ is the Lambert's W function defined implicitly by the
equation $z=W(z)e^{W(z)}$.  Note that, by using an expansion
$W(z)=\log z -\log \left(\log z\right) +\cdots$ for large $z$, we
obtain
\begin{align} 
\Delta_0=\sqrt{\frac{\tilde{\lambda'}}{2\pi}\log\left(
\frac{2\pi \Lambda^2}{\tilde{\lambda'}} \right) }
+\cdots.
\label{delta-0}
\end{align} 
Imposing the renormalization condition (\ref{choice}),  we then obtain
a relation\footnote{\label{ftnt-choice2} The inequality follows from
  the condition $\tilde{\lambda}/M_{KK}^2 \ll 1 $, which is necessary
  for a KK reduction. This condition implies that the dimensionless 'tHooft
  coupling is small: $\tilde{\lambda}_{D+2} M_{KK}^{D-2} \ll 1$.}
\begin{align} 
\Delta_0^2/\tilde{\lambda} \sim \frac{1}{2\pi}\log\left(
\frac{2\pi M_{KK}^2}{\tilde{\lambda}} \right) \gg 1 
\label{choice2}.
\end{align} 
Thus we confirmed that the assertion (a) above is satisfied.
Substituting \eq{delta-value}  in $S_{eff}$ in \eq{s-eff} and ignoring the
term involving Tr$V$, we get the following effective action for $A_\mu $:
\begin{align} 
\frac{S(A)}{D N^2}&= C(\tilde \l', \Delta_0) + \int_0^\beta \kern-5pt
dt\, \int_0^L \kern-5pt dx \left(\frac{1}{2\tilde \l N} {F_{01}^a}^2 -
\frac{1}{\sqrt{2\pi}} \sqrt{\frac{\Delta_0}{\beta^3} }
e^{-\Delta_0 \beta } \left| \frac{1}{N} \Tr U \right|^2 \right) +
\cdots,
\label{Seff-V} \\
&C(\tilde\l', \Delta_0)= \frac{\beta L \Delta_0^2}{8\pi}\left(1+
\frac{\pi \Delta_0^2}{\tilde{\lambda'} }   \right),
\end{align}
where the terms $\cdots$ are higher order in the same sense as in
\eq{only-u}. We have used $\int dx dt |\Tr U|^2 = L\b |\Tr U|^2 $
which is correct up to derivatives of $U$ which occur at $O(\tilde
\l/\Delta^2)$ and are small, as argued above.

In the limit of $L \to \infty$\footnote{Although we are apparently
  taking $L \to \infty$ here, as we discuss at the end of the next
  subsection, we can extend these results to finite $L$ which is large
  enough to ensure vanishing ot Tr$V$.  Note that if $\Tr V \neq 0$,
  the $1/D$ expansion does not work at $L \to \infty$ as argued in
  \cite{Morita:2010vi}, where such a situation is discussed in the
  presence of $R$-symmetry chemical potential.} we can choose the
gauge $A_1 =0$. Solving the Gauss's law condition in this gauge, we
get $A_0= (g^2/\del_x^2) \varrho$ where $\varrho \equiv -
\frac{i}{2}[Y^I,\del_t Y^I]$ is the charge density; in the large $D$
saddle point, especially for large enough $\Delta_0$, the condensate
is static and temporal fluctuations of $\varrho$, and consequently, of
$A_0$, are suppressed. As a result, we can write
\[ \int dx\ dt\ \Tr F_{01}^2 = \b \int dx \Tr( \del_x A_0)^2=
\b^{-1} \int dx \Tr |\del_x U|^2,
\]
where
\begin{align}
U(x) = P \exp[i\int_0^\b dt A_0(x,t)]= \exp[i\b A_0(x)]
\label{u-x}.
\end{align} 
By rescaling $ x \to x'= \tilde{\lambda} \beta x$ we eventually get an
effective action in terms of $U(x)$
\begin{align}
S/DN^2= C(\tilde \l', \Delta_0) + \int_{-\infty}^\infty\kern-5pt dx 
\left[\frac{1}{2N} \Tr \left(| \del_x U|^2 \right) -\frac{\xi}{N^2} 
  \left|\Tr U\right|^2 \right], 
\label{eff-action}
\end{align} 
where
\begin{align} 
\xi= \sqrt{\frac{\Delta_0}{2\pi \tilde{\lambda}^2
    \beta^3}}e^{-\Delta_0 \beta}.
\label{def xi}
\end{align}
Note that $\xi$ is a monotonically decreasing function of $\beta$. 

Equations similar to \eq{eff-action}, \eq{def xi} have earlier been
derived in \cite{Semenoff:1996xg} who consider a two-dimensional gauge
theory with heavy adjoint scalars of mass $m$. In their equations $m$
appears in place of $\Delta_0$. 
We could have, in fact, derived the
above effective action \eq{eff-action} as follows: the large $D$
saddle point generates a dynamical mass $\Delta_0$ for the adjoint
scalars which turns the theory into a massive adjoint scalar QCD; once
this is established, we can use the method of \cite{Semenoff:1996xg}
with $m = \Delta_0$, to arrive at \eq{eff-action}. The agreement with
the results of \cite{Semenoff:1996xg} provides an additional check on
our derivation. 
Adjoint scalar QCD with large scalar mass has also
been considered by \cite{Aharony:2005ew} who have independently
derived equations analogous to \eq{eff-action}, \eq{def xi}; our
method of derivation follows their derivation closely, except that our
mass is dynamically generated, as mentioned above.  The discussion in
appendix \ref{app-massive} involving arbitrary $m$ relates the two
extreme cases of large mass and zero mass.

\subsection{The phase transition at Large $L$}
\label{sec-phase tr}

Phase transitions in the system \eq{eff-action} have been discussed in
\cite{Semenoff:1996xg, Gattringer:1996fi, Basu:2006mq}. We will adopt
their result to infer about phase transitions in our two-dimensional
gauge theory \eq{d=2} at large $L$ (the range of $L$ is defined in
\eq{large-L}). For completeness, we will briefly review some of the
results in these papers.

If we regard the coordinate $x$ in \eq{eff-action} as time, it becomes
the quantum mechanics of a single unitary matrix, a subject that has
been extensively studied \cite{Wadia:1980cp},
\cite{Sengupta:1990bt}-\cite{Dhar:1992hr}. Phases of such a model can
be described by the behaviour of the eigenvalue density
\begin{align}
\rho(\theta,x) =  \frac1{N} \sum_{i=1}^N \delta(\theta - \theta_i(x))
\label{eigen-density},
\end{align}
where $\exp\left( i\theta_i(x)\right) $ are the eigenvalues of 
\eq{u-x}.

The hamiltonian of this system (regarding $x$ as time) can be written
as \cite{Sengupta:1990bt}-\cite{Dhar:1992hr} 
\begin{align} 
H=\int d \theta \left( \frac{1}{2}\rho v^2+\frac{\pi^2 }{6} \rho^3 \right)  
-\xi \left| u_1 \right|^2 
-\frac{1}{24} ,
\label{collective-ham}
\end{align} 
where we have ignored the constant term $C$ in \eq{eff-action}.
Here $v= \del_\theta \Pi$, and $\Pi(\theta,x)$ is the canonical
conjugate of $\rho(\theta,x)$. We have also used the notation
$u_n = \frac1N \Tr U^n, u_{-n} = (u_n)^*$. Note that 
$u_n(x)$ are moments of the eigenvalue density \eq{eigen-density}:
\begin{align} 
\rho(\theta,x)
&= \frac{1}{2\pi}  \sum_{n=-\infty}^\infty u_n(x) e^{-in \theta}
\label{moments}.
\end{align} 
To study the various equilibrium phases of the system, we study static
solutions of (\ref{collective-ham}), which \cite{Semenoff:1996xg,
  Gattringer:1996fi, Basu:2006mq} are given by $v(\theta)=0$ and
\begin{align} 
\rho(\theta)=\frac{\sqrt 2}{\pi} \left(\sqrt{E+2\xi \rho_1 \cos \theta}
\right),
\label{static-soln}
\end{align} 
where $\rho_1 = u_1 = u^*_1$ is the first moment (real in this case),
which must self-consistently satisfy (see \eq{moments})
\begin{align}
\int_0^{2\pi}d\theta \rho(\theta) \cos\theta = \rho_1.
\label{consistency}
\end{align}
The constant $E$ is fixed by solving the normalization condition
\begin{align}
\int_0^{2\pi} d\theta\, \rho(\theta) =1.
\label{normalization}
\end{align}
Depending on the value of the constant $\xi$ in \eq{collective-ham},
Eqn. \eq{normalization} may not determine $E$ uniquely. In general, we
obtain three branches $E(\xi)$, depending on whether \eq{static-soln}
describes a uniform, non-uniform or gapped eigenvalue distribution
(see figure \ref{fig-3-phase}).
\begin{figure}[H]
\begin{center}
\includegraphics[scale=.65]{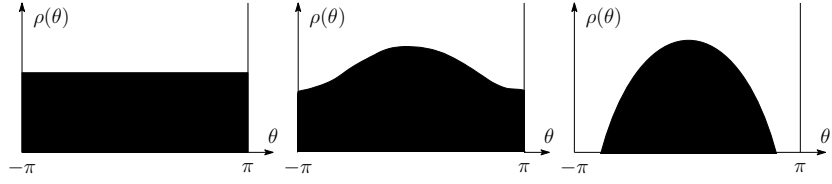}
\caption{Configurations of eigenvalue density $\rho(\theta)$ in the
  unitary matrix model.  The left plot is the uniform distribution
  (corresponding to $\rho_1=0$ in \eq{static-soln}), the middle one is
  the non-uniform distribution ($|E/(2\xi \rho_1)| \ge 1$) and the
  right one is the gapped distribution ($|E/(2\xi \rho_1)| \le 1$).}
\label{fig-3-phase}
\end{center}
\end{figure}
The value of $E(\xi)$ in these three cases are different.  The
thermodynamic stability for the various branches of the function
$E(\xi)$ is analyzed \cite{Semenoff:1996xg, Gattringer:1996fi,
  Basu:2006mq} by comparing the values of the Euclidean Hamiltonian
\eq{collective-ham} which can also be regarded as the free energy.
They can be summarised as follows:
\begin{itemize}
\item Independent of the value of $\xi$, the uniform solution always
  exists. We call this phase as I.
\item For $\xi< \xi_0=0.227$, only one solution (phase I) exists and
  is stable.
\item At $\xi=\xi_0$, there is nucleation of two gapped solutions. One
  is unstable (phase II) and another is meta-stable (phase III).
\item At $\xi=\xi_1=0.23125$, a GWW type phase transition \cite{Gross:1980he, Wadia:1980cp} occurs in
  phase II and the gapped solution becomes a solution with non-uniform
  distribution (Phase IV).
\item At $\xi=\xi_2=0.237$, there is a first order phase transition
  between the phases I and III. 
Above $\xi_2$, the phase III is stable and the phase I is meta-stable.
\item At $\xi=\xi_3=1/4$,  phase IV merges into phase I, and
  the uniform solution becomes unstable beyond $\xi_3$.
\end{itemize}
These are summarised in figure \ref{fig-potential}.

\begin{figure}[H]
\begin{center}
\includegraphics[scale=.35]{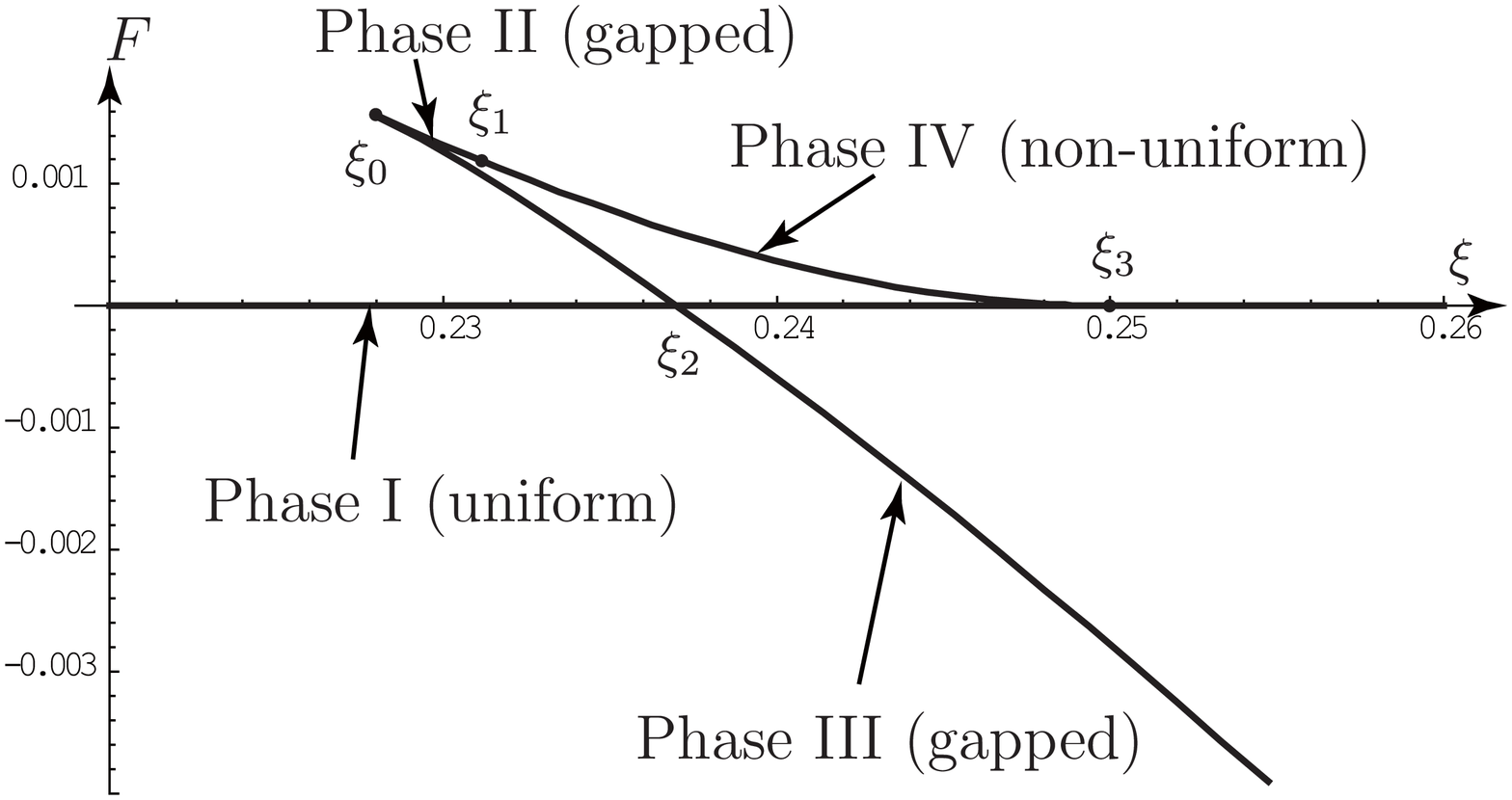}
\caption{
Free energy (\ref{collective-ham}) vs $\xi$ in the four phases.
The gapped and non-uniform solutions here are numerically evaluated.
  Since $\xi$ is
  a monotonically increasing function of temperature (see \eq{def
    xi}), the uniform distribution (Phase I) is stable at low
  temperatures and the gapped distribution (Phase III) is stable at
  higher temperature.  A first order phase transition between these
  two phases happens at $\xi_2$.}
\label{fig-potential}
\end{center}
\end{figure}
\gap{-2}
Using (\ref{def xi}), we can read off the critical temperatures
corresponding to these transition points:
\begin{align} 
\beta_m &\equiv \beta(\xi_m)=
\frac{3}{2\Delta_0}W\left[ \frac{2}{3(2\pi \xi_m^2)^{1/3}}
  \left(\frac{\Delta_0^2}{\tilde{\lambda}} \right)^{2/3} \right]
\nn 
\approx & \frac{1}{\Delta_0}\left( \log\left(
\frac{\Delta^2_0}{\tilde \lambda}\right) -1.53 - \log
\xi_m\right),\quad m=0,1,2,3.
\label{temperature}
\end{align}
Here Lambert's W function is employed again. In the second step we
have assumed $\Delta_0^2/\tilde {\lambda} \gg 1$ and $\xi_m = O(1)$.

As we come down from $\beta=\infty$ (go up in temperature), there is a
first order phase transition at $\b_2$ from the centre symmetric phase
($\Tr U=0$) to the broken symmetry phase ($\Tr U \ne 0$) at an inverse
temperature
\begin{align}
\b_{cr} \equiv \b_2 \approx \frac{1}{\Delta_0} \log\left(
\frac{\Delta^2_0}{\tilde \lambda}\right) .
\label{beta-2}
\end{align}
Several comments are in order here:
\begin{itemize}

\item[(a)] The assumption, used in Section \ref{sec-large-D}, that
  $e^{-\Delta \beta}$ is small, is correct in the parameter region we
  are interested in.  The interesting phase structures appear in the
  regime $\xi\sim O(1)$.  Thus since $\tilde{\lambda}/\Delta_0^2 \ll
  1$ (see \eq{choice2}), $e^{-\Delta \beta} \sim e^{-\Delta_0 \beta}
  \ll 1$  from (\ref{temperature}).  Therefore the terms in
  (\ref{arbit-u-v}) involving $U^n, n=2,3,...$ are suppressed.

\item[(b)] The Euclidean model \eq{d=2} is symmetric under the
  interchange of $(t,\b) \leftrightarrow (x, L)$. Hence, similarly to
  \eq{beta-2}, we can deduce a phase transition in $L$ from the $\Tr
  V=0$ phase to $\Tr V \ne 0$ (at large enough $\b$) at a critical
  length $L_{cr}=\beta_{cr}$.

\item[(c)] The existence of a finite $L_{cr}$ above confirms that the
  transition which we found at $L\to \infty$ and $\b= \b_{cr} $
  between $\Tr U=0$ and $\Tr U \neq 0$ indeed happens in the $\Tr V=0$
  phase.  Therefore the expression for $\beta_{cr} $ is valid even at
  finite $L$ as long as $\Tr V =0$, since large $N$ volume
  independence \cite{Eguchi:1982nm, Gocksch:1982en} ensures that gauge invariant quantities like the free energy and 
vev of Wilson loop operators do not depend on
  $L$ in the $\Tr V =0$ phase.
   Thus, the correct definition of `large $L$' in this section is
  \begin{align}
    L \gg  L_{cr} =\beta_{cr},
    \label{large-L}
  \end{align}
  which ensures that we are in the $\Tr V=0$ phase. $\b_{cr}$ is
  defined in \eq{beta-2}.

\item[(d)] By considering the interchange $\beta \leftrightarrow L$,
  we can claim that, if there is no direct transition from $\Tr U =
  \Tr V =0$ phase to $\Tr U \neq 0$ $\Tr V \neq 0$ phase, the two transition line $\beta = \beta_{cr}$ and $L
  = L_{cr}$ meet at $\beta=L=\beta_{cr}$. See figure \ref{fig-phases}.\footnote{Note
    that a transition line between $\Tr U = \Tr V =0$ phase and $\Tr U
    \neq 0$ $\Tr V \neq 0$ phase, if it exists (e.g. as in the third
    circled option in figure \ref{fig-phases}), could depend on $L$
    and $\beta$.  However the gravity analysis in Section
    \ref{sec-gravity} suggests that there is no such transition in our
    model, consistent with the `cascade picture' reviewed in Section
    \ref{other}. In this case, as represented by the second joining option
in figure  \ref{fig-phases}, \eq{large-L} can be relaxed to $L> L_{cr}$.}

\item[(e)] As we discuss in appendix \ref{app-massive}, the non-zero
  mass in (\ref{d=2}) does not change the qualitative nature of the
  phase structure.
\end{itemize}

\section{\label{sec-small-L} Phase transitions at small $L$}

In this section, we discuss the phase structure for small $L$ ($L \ll
L_{cr}$).  In this case we can dimensionally reduce the theory (see
footnote \ref{tricky}) to obtain the action (\ref{d=d}) with $d=1$.
Hence we can use the analysis in \cite{Mandal:2009vz}
\footnote{In \cite{Mandal:2009vz} we had considered massless adjoint
  scalars.  We generalize the results to non-zero mass in Appendix
  \ref{app-massive}.} where the phase structure has been studied by
using the $1/D$ expansion.  The phases are characterized by the
eigenvalue density (\ref{moments}). Here we summarise the results of
\cite{Mandal:2009vz}, 
\begin{itemize}
  \item $\beta > \beta_{c1}$: The stable solution is given by $u_n=0$
    ($n \ge 1$). The eigenvalues of $A_0$ are distributed uniformly.
  \item $ \beta_{c1} > \beta >\beta_{c2}$: 
  The stable solution is given by $u_1 \ne 0 $, $u_n=0$ ($ n \ge 2 $).
  The eigenvalue distribution is non-uniform and gapless.
  \item $\beta_{c2} > \beta$: The stable solution is given by $u_n \ne
    0$ ($n \ge 1$).  The eigenvalue distribution is gapped.
  \item The phase transition at $\beta = \beta_{c1}$ is of 
    second order and the transition at $\beta = \beta_{c2}$ is a
    third order (GWW type) transition.
\end{itemize}
The critical temperatures are calculated up to $O(1/D)$ in
\cite{Mandal:2009vz} \footnote{Eqns. \eq{tc1},\eq{tc2} are calculated
  for $m=0$.  The massive case is discussed in appendix
  \ref{app-massive}.  Note that the mass from the KK modes for $d=1$
  is proportional to $\sqrt{\lambda_1 L} $ as in
  (\ref{m-ren-1}). Thus the mass correction is small for small
  $L$.} as
\begin{align} 
\beta_{c1} \tilde{\lambda}^{1/3}_1 &  = \log \tilde{D}
\left( 1 +\frac{1}{\tilde{D}} \left( \frac{203}{160} -\frac{\sqrt{5}}{3}
\right) \right) ,
\label{tc1}\\
\beta_{c2} \tilde{\lambda}^{1/3}_1
&-\beta_{c1} \tilde{\lambda}^{1/3}_1 \nonumber \\
 =&\frac{\log \tilde{D}}{\tilde{D}} \Biggl[ -\frac{1}{6}    
 +\frac{1}{\tilde{D}}
 \left( 
 \left( -\frac{499073}{460800}+\frac{203\sqrt{5}}{480}   
 \right) \log \tilde{D} 
 -\frac{1127\sqrt{5} }{1800}+\frac{85051}{76800} 
  \right) 
  \Biggr],
  \label{tc2}
\end{align} 
where $\tilde{D}=D+1$ and $\tilde{\lambda}_1=(g')^2 N (D+1)/L $.  In
the $\b$-$L$ plane the transition lines appear as curves $\b \propto
L^{1/3}$ passing through the origin. Since our analysis is valid only
for $L \ll L_{cr}$, we should trust these transition lines only in
that region, as we have depicted in figure \ref{fig-phases}.  By using
the $\b \leftrightarrow L$ reflection symmetry, we can also infer
phase transition lines for $\b \ll \b_{cr}$ described by $L \propto
\b^{1/3}$, as shown in figure \ref{fig-phases}.

Contrary to the case of large $L$, an intermediate non-uniform phase
exists at small $L$.  A similar feature in the context of higher order
confinement/deconfinement type phase transitions has been seen in
\cite{Aharony:2003sx}.

We should mention that considerations in this
section are valid  up to $\tilde{\lambda}' \lesssim \lambda_{max}$ where
$\lambda_{max}= L/\b^3$ for $\b \ll \b_{cr}$, and $\lambda_{max}=\b/L^3$ for
$L \ll L_{cr}$ \cite{Mandal:2009vz}, which can be large close to the
origin (See footnote \ref{ftnt-d1-strong} also.).
We will come back to this point in
Section \ref{sec-gravity}.

\section{\label{sec-phase-gauge} Phases of 2d gauge theory on $T^2$}

In the last two sections, we have studied confinement/deconfinement
type transitions in the model \eq{d=2} for large and small values of
the spatial size $L$.  

We have found that the nature of the transition depends on $L$.  
We can summarise these results in
figure \ref{fig-phases}, where we supplemented our calculations with
the reflection symmetry $\beta \leftrightarrow L$ of the model.
\begin{figure}[H]
\begin{center}
\includegraphics[scale=0.8]{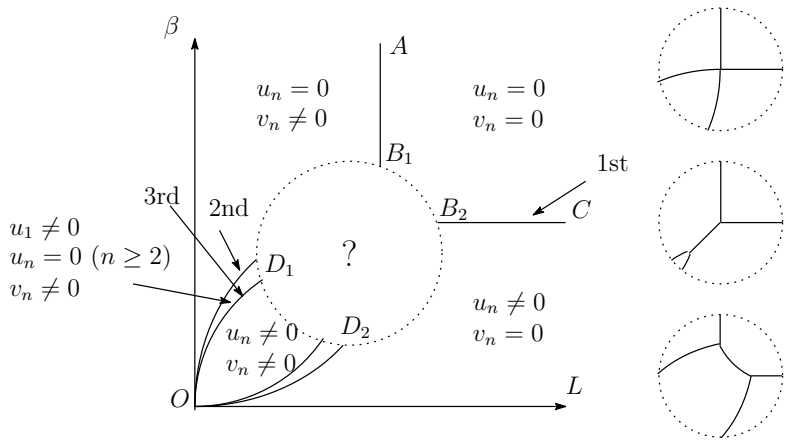}
\caption{Phase structure of the 2d gauge theory at weak coupling
  (defined below). There are essentially 4 phases characterized by
  non-zero values of various Wilson lines. The inner region, with both
  Wilson lines non-zero, includes 2 additional phases in which the
  eigenvalue distribution is gapless but non-uniform.
The orders of the phase transitions (1st, 2nd, 3rd) are indicated.  Our analysis does 
not apply to the region enclosed by the dotted lines.
Possible connections between the phase boundaries across this region are suggested in the inset (where
  boundaries of the intermediate phases are omitted for simplicity).
  A similar diagram is proposed in \cite{Aharony:2005ew} for the model
  (\ref{d=2}) with large mass for the adjoint scalars (see Section
\ref{other-analytic} for details).  As we will see
  in figure \ref{fig-phases-gr}, the gravity analysis conforms to the
  second pattern. We will see in section \ref{lattice} that the second
  pattern is also supported by lattice studies.}
\label{fig-phases}
\end{center}
\end{figure}

{\em Weak coupling} in the above diagram (figure \ref{fig-phases}),
for large $L$ (or large $\beta$), is
defined by $\tilde{\lambda}/ \Delta^2 \ll 1 $ (see assumption (a)
below \eq{logdet}). In case the 2d gauge theory is obtained from a KK
reduction, the above notion of weak coupling translates to $\tilde\l
\ll M_{KK}^2$ (see \eq{choice2} and footnote \ref{ren-cond}).
For small $L$ (or $\beta$), the coupling should satisfy $\tilde{\lambda}' \lesssim \beta/L^3 $ (or $L/\beta^3$) to
validate the additional KK reduction to one dimension; as remarked
at the end of Section \ref{sec-small-L}, this limit on
$\tilde{\lambda}'$ can be quite large close
to the origin of figure  \ref{fig-phases}.

As mentioned in the Introduction, our model \eq{d=2} can be regarded
as a dimensional reduction of a $D+2$ dimensional pure Yang-Mills
theory compactified on a small $T^D$. For the dimensional reduction to
work (see footnote \ref{tricky}), $W_I= \Tr U_I$ ($I=d,\cdots,d+D-1$)
must be non-zero (which is ensured by a sufficiently small size of the
$D$-dimensional torus). Therefore we can regard the phase structure in
figure \ref{fig-phases} as a part of the $D+2$ dimensional pure
Yang-Mills theory in the $W_I \neq 0$ phase.  Such a Yang-Mills theory
on $T^3$ and $T^4$ have been studied in lattice gauge theory and we
will compare our results with those studies in Section \ref{other}.

Since our phase structure is derived through the $1/D$ expansion, it
is not {\em a priori} obvious whether the result should be valid for
small $D$.  However, at least for small $L$, the comparison with
numerical studies \cite{Kawahara:2007fn, Azuma:2007fj,
  Azeyanagi:2009zf}, as explained in \cite{Mandal:2009vz}, turns out
to be remarkably good even for small $D$. For example, for $D=2$ the
$1/D$ expansion, performed up to an accuracy of $O(1/D)^2$ reproduces
numerical results within the expected 25\%. Thus we believe that the
phase structure in the large $L$ region also should be qualitatively
correct for small $D$ ($D \ge 2$).

\section{The phase structure from gravity}
\label{sec-gravity}

In the previous sections, we evaluated the phase structure of the
bosonic gauge theory (\ref{d=2}) at weak coupling.  In this analysis,
it was difficult to figure out the phase structure of the middle
region, namely where both $\beta$ and $L$ have intermediate values. In
particular, it was not clear how the various phase boundaries in
figure \ref{fig-phases} are connected.

In this section, we attempt to construct a gravitational dual of our
system along the lines of Witten's realization of the 4d pure
Yang-Mills theory \cite{Witten:1998zw}.  We consider IIA supergravity
on $R^7 \times T^3$ ($T^3=S^1_\beta \times S^1_{L}\times S^1_{L_2}$)
and put $N$ D2 branes on the $T^3$. The AdS/CFT duality in this
context is discussed in \cite{Itzhaki:1998dd} and more details are
provided in Appendix \ref{app-grav}. The dual gauge theory is 3
dimensional $\mathcal{N}=8$ $SU(N)$ super Yang-Mills on $T^3$, with
the identifications $(t,x_1,x_2)=(t+\beta,x_1+L,x_2+L_2)$.  In order
to complete the definition of the theory, we need to choose a boundary
condition of the fermions along each compact direction.  Let us impose
an anti-periodic boundary condition on the fermions on the $x_2$
cycle.  If $L_2$ is sufficiently small ($1/L_2 \gg \lambda_3 = g_3^2
N$ \footnote{This implies $1/L_2 \gg \sqrt{\l_2}$.}), we can use a
dimensional reduction to the two-dimensional torus $S^1_\beta \times
S^1_{L}$: owing to the anti-periodic boundary condition along $L_2$,
all fermions would acquire a mass proportional to $1/L_2$ and we can
ignore them\footnote{\label{ftnt-mass} In fact, even the scalars would
  acquire a mass at one-loop, as in \cite{Witten:1998zw}.  However,
  unlike in \cite{Witten:1998zw}, the scalar mass does not become
  infinite as $L_2\to 0$. From the 2d gauge theory perspective, the
  scalar mass renormalization due to fermion loops schematically goes
  as $ m_Y^2= g_2^2 N \int d^2 p \frac{1}{(m) ( p + m) } = \lambda_2
  \frac{1}{m} \Lambda = \lambda_2.$ where we have used a fermion mass
  $m= 1/L_2$ and an uv cut-off for the 2d theory $\Lambda= 1/L_2$.
  For a more precise calculation see appendix \ref{KK loop}. Since the
  scalars remain light compared to the KK scale, we must keep them in
  the Lagrangian as in \eq{d=2}.}.

One would, thus, expect the gravitational system, for small
$L_2$, to describe a dual of \eq{d=2} with $D=8$.  Unfortunately,
however, the gravity solutions are not valid in the small $L_2$ region
($L_2 \ll 1/\lambda_3$) since stringy corrections become important
(see appendix
\ref{sec-gr-valid}).\footnote{ \label{ftnt-strong-coupling} This is a
  common problem in the construction of holographic duals of
  non-supersymmetric gauge theories. Since the gauge theory coupling
  constant $\lambda_3$ is greater than the KK scale ($1/L_2$) in the
  region of validity of gravity, the gravity description has been
  likened (cf. \cite{Aharony:1999ti}, p. 196-197) to strong coupling
  lattice gauge theory, the small $L_2$ limit being regarded as
  analogous to the continuum limit. Interesting results, including the
  qualitative predictions in \cite{Witten:1998zw}, and results in
  AdS/QCD, have been obtained using this philosophy. We will use the
  small $L_2$ extrapolation of our gravity results in this spirit.}
In case of $L_2 \gg 1/\l_3$, since the fermions are not decoupled, the
D2 brane theory will depend on the boundary conditions of fermions
along the $t$ and $x_1$ directions.  There are 4 choices of boundary
conditions: (AP,AP), (AP,P), (P,AP), (P,P), where P denotes the
periodic boundary condition and AP denotes the anti-periodic
one\footnote{The gravity calculation with $\beta=L$ in the (P,P) case
  has been studied in \cite{Martinec:1998ja, Hanada:2007wn}. }.  
 Phase diagrams of the gravity theories for different spin structures are
worked out in Appendix \ref{app-grav} and presented in figures
\ref{fig-phases-gr}, \ref{fig-APAPAP-L} and \ref{fig-APPAP-L}. The salient features are:

(i) The phase structures in the gravity analysis depend on the boundary conditions.

(ii) Only the gravity analysis with (P,P) boundary condition is reliable as a prediction for  gauge theory through the arguments in appendix \ref{app-conjecture}. The phase structure in this case is shown in figure \ref{fig-phases-gr}.
It predicts the second joining pattern in figure \ref{fig-phases}.

(iii) The phase transitions in the gravity description in figure \ref{fig-phases-gr} are
Gregory-Laflamme (type) transitions \cite{Gregory:1994bj, Kol:2004ww,
  Harmark:2007md} and are expected to be of the first order, at least for large $L_2$ \cite{Aharony:2004ig, Hanada:2007wn}.

(iv) The gravity analysis for small $\beta$ and $L$ is not reliable.

\begin{figure}[H]
\begin{center}
\includegraphics[scale=.8]{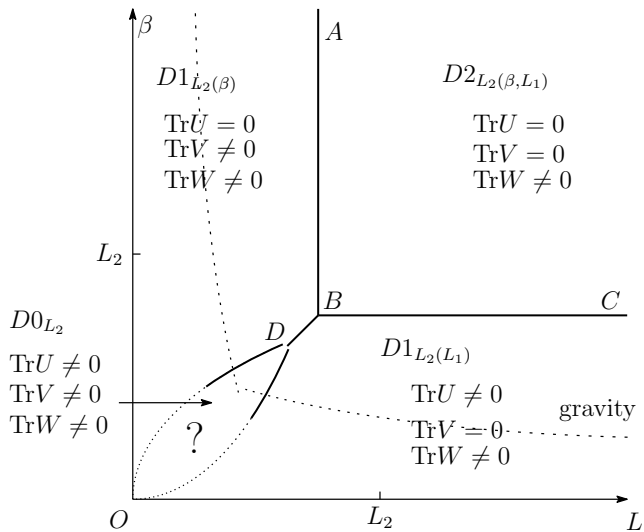}
\caption{Conjectured phase structure of the gauge theory from the gravity analysis. 
 We used a large $L_2$ and the (P,P,AP) spin structure of the fermions on the three-dimensional torus (with AP boundary condition on the Scherk-Schwarz (SS) circle). Tr$W$ is the Polyakov loop operator along the SS circle (\ref{Polyakov3}).
  The gravity analysis is reliable only in the region above the dotted line.
The transitions in this diagram are predicted to be first order phase transitions.
(See appendix \ref{app-conjecture})
}
\label{fig-phases-gr}
\end{center}
\end{figure}

Comparing figures \ref{fig-phases} and \ref{fig-phases-gr}, we can see
that both diagrams share some common features.  Both have four phases
in similar parameter regions. In particular, the behaviour of the
transition lines for large $\beta$ and large $L$ is the same.  The
line $BC$ in figure \ref{fig-phases-gr} is independent of $L$. This is
consistent with large $N$ volume independence \cite{Eguchi:1982nm,
  Gocksch:1982en,Poppitz:2010bt}, since $\Tr V= 0$ on both sides of $BC$.  Similar
remarks apply to the line $AB$ as well.

In the small $\beta, L$ region too the two phase diagrams share
similarities. In figure \ref{fig-phases-gr}
two phase transition lines emanate from the point $D$ towards low
values of $\b, L$.
However it is not clear from the gravity analysis how to
continue towards the origin.
On the other hand, in figure \ref{fig-phases} the region near the origin $O$ can be computed reliably and the two (double) lines $OD_1$ and $OD_2$ can be identified as a continuation of the two phase
transition lines mentioned above.
We should note that the phase structure in the small $\b, L$ region of
figure \ref{fig-phases} can be calculated from gauge theory even at
strong coupling, up to $\tilde{\lambda}' \lesssim \lambda_{max}$ where
$\lambda_{max}= L/\b^3$ for $\b \ll \b_{cr}$, and $\lambda_{max}=\b/L^3$ for
$L \ll L_{cr}$; as $\l$ grows stronger, the calculable region becomes
narrower.

In addition to the above similarities, various details of the phases
in figure \ref{fig-phases} and \ref{fig-phases-gr} are also similar.
Recall that the phases in the gauge theory are characterized by three
solutions: uniform, non-uniform and gapped.  Correspondingly, three
solutions (uniform black string, non-uniform black string and localized
black hole) play a key role in the discussion of Gregory-Laflamme
transitions in gravity.  For large $L$, the free energy of these
solutions in the gauge theory are related as shown in figure
\ref{fig-potential}.  A similar relation has been found in gravity
\cite{Kudoh:2004hs}, in the case where the GL transition is of first
order.  On the other hand, if the GL transitions are of higher order,
it consists of two transitions: a transition between a uniform black
string and a non-uniform one, and another transition between the
non-uniform black string and a localized black hole \cite{Kol:2004ww}.
This is precisely similar to the higher order phase transitions in the
gauge theory, which we have observed in the small $L$ case.

An important consequence of the gravity analysis is that we can guess
how the phase transitions in figure \ref{fig-phases} are connected.
In particular, it indicates that there is no direct transition between
$\Tr U= \Tr V=0$ and $\Tr U \neq0, \Tr V \neq 0$ phase.  In
\cite{Hanada:2007wn}, it was pointed out that this property has also
been observed in large $N$ pure Yang-Mills theories on 4 and 3
dimensional tori in the lattice calculations of
\cite{Narayanan:2003fc, Kiskis:2003rd, Narayanan:2007ug}. Thus the
gravity analysis is in agreement with the lattice calculation. More
details of the lattice calculation are presented in Section
\ref{other}. 

In summary, the full phase diagram of the two dimensional gauge theory
(\ref{d=2}) may be obtained by combining the results from gauge theory
and gravity. The result would be given by figure \ref{fig-phases} with
the second joining possibility.
\footnote{\label{support}The extrapolation involved in this conclusion has
additional support from the lattice calculations mentioned above.}

\section{\label{other}  Relation to other works}

In this section, we detail some of the remarks made in the
Introduction regarding previous works.

\subsection{\label{lattice}Comparison with lattice studies}

Large $N$ Yang-Mills theories on tori have been studied
using lattice methods in \cite{Narayanan:2003fc, Kiskis:2003rd,
  Narayanan:2007ug, Narayanan:2007fb}, in  $d=3$ and 4 dimensions.
Reference \cite{Narayanan:2007fb} contains a nice summary of these works.
We describe some of the salient features below (see also figure \ref{fig-lattice}).

(a) If we start from $d=4$ pure Yang Mills theory on a Euclidean torus
$T^4$ with $L_3 < L_2 < L_1 < L_0$ \footnote{Since all directions are
  equivalent in the Euclidean space, the ordering chosen here is
  arbitrary and all arguments below can be repeated with any other
  ordering.}, then for all radii large enough the centre symmetry
$Z_N^4$ (see footnote \ref{ftnt-centre}) is unbroken and all the
Wilson loops $W_\mu$ vanish. This phase is called the $0_c$ phase. In
this completely unbroken phase, the thermodynamics in the large $N$
limit does not depend on any one of the lengths $L_\mu$
\cite{Eguchi:1982nm, Gocksch:1982en, Unsal:2010qh}.

(b) As $L_3$ is decreased, below a certain value $L_3^c$ there is a
phase transition to a new phase where the centre symmetry is broken to
$Z_N^3$ and $W_3$ develops a non-zero expectation value. The other
Wilson loops $W_0, W_1$ and $W_2$ still vanish. This phase is called
the $1_c$ phase in which there is no dependence on the lengths $L_\mu,
\mu=0,1,2$ which characterize the directions of unbroken centre
symmetry.

(c) If $L_2$ is now decreased, maintaining $L_2 > L_3$, a new phase
$2_c$ appears below $L_2^c$ (which is a function of $L_3$) where $W_2$
becomes non-zero.  The centre symmetry is broken to $Z_N^2$, with
non-zero values of $W_2, W_3$ while $W_0, W_1$ still vanish.

(d) Proceeding similarly, a phase $3_c$ is reached when $L_1$ is
reduced below a critical value $L_1^c(L_2, L_3)$, and the phase $4_c$
are reached when, finally, $L_0$ is reduced below $L_0^c(L_1, L_2,
L_3)$.

(e) Thus, $d=4$ pure YM theory (with $L_3< L_2 < L_1 < L_0$) exhibits
a cascade of transitions
\[
0_c \to 1_c \to 2_c \to 3_c \to 4_c.
\]

(f) In case of $d=3$ pure Yang Mills theory (with $L_2 < L_1< L_0$)
the sequence of transitions works similarly, leading to a cascade
\[
0_c \to 1_c \to 2_c \to 3_c.
\]

(g) It was found in \cite{Narayanan:2003fc, Kiskis:2003rd} that the
cascade of transitions persists even when all radii are the
same. E.g. in case of the system in (f) with $L_1=L_2=L_3=L$, for high
enough $L$ all $W_\mu=0$; as $L$ is reduced below a certain critical
value $L_c$, only one of the $W_\mu$'s picks up a non-zero value
\cite{Narayanan:2003fc}, and the 3D cubic symmetry group spontaneously
breaks down to the symmetry group appropriate to a square lattice.

(h) Generally speaking, it was found in these works that the Wilson
lines $W_\mu$ change from zero to non-zero one by one; two or more
Wilson lines never simultaneously change from zero to non-zero values.

(i) {\em Order of phase transitions:}\footnote{We
thank Rajamani Narayanan for pointing out some references in
this paragraph.} There is ample evidence that the
first of the cascade of transitions $Z_N^{d+D} \to Z_N^{d+D-1}$ is
first order.  In the case of $0_c \to 1_c$ transitions in
four-dimensional Yang Mills theories on $T^4$ such evidences are
presented directly in \cite{Kiskis:2005hf} and indirectly, assuming
large $N$ volume independence, in \cite{Lucini:2003zr, Panero:2009tv,
  Datta:2010sq}. Evidences for the first order nature of the $0_c \to
1_c$ and $1_c \to 2_c$ phase transitions have been presented for Yang
Mills theory on $T^3$ in \cite{Koren2009}, which indicates that the
first two transitions $Z_N^{d+D} \to Z_N^{d+D-1} \to Z_N^{d+D-2}$ are
also first order. Our gauge theory analysis presents analytic evidence
that the first order nature continues till the transition $Z_N^{2} \to
Z_N^{1}$ (which is $2_c \to 3_c$ in the notation of pure Yang Mills
theory on $T^4$), whereas the last transition $Z_N \to 1$ in which the
centre symmetry is completely broken, occurs (in the parameter region of Section \ref{sec-small-L} \footnote{In other parameter regions $3_c \to 4_c$ can be a single first order phase transition\cite{Mandal:2009vz, Koren2009}.})  in two steps through a
second and a third order phase transitions. The last statement, first
derived in \cite{Mandal:2009vz}, is corroborated by the numerical work
in \cite{Kawahara:2007fn}.

Let us compare the above with the phase diagram in figure \ref{fig-phases}, which describes phases of the theory \eq{d=2}. Note
that the theory \eq{d=2} with $D=2$ \footnote{We will assume that the
  essential features of the large $D$ calculation which led to this
  phase diagram remain valid for $D=2$. This was certainly true in the
  $d=1$ case discussed in \cite{Mandal:2009vz}.} is precisely the one
obtained after the steps (a)-(b) described above, in which we reduce a
$d=4, D=0$ theory (pure YM theory in four dimensions) on two small
circles of length $L_2, L_3$. To make the correspondence more
explicit, we identify $L_1 = L, L_0= \b$, $W_1 = {\rm Tr} V, W_0 =
{\rm Tr} U$. It is easy to see that the phase transitions in the $L <
\b$ region of figure \ref{fig-phases} precisely correspond to the phase
transitions described in (c)-(d) above\footnote{The $\b < L$ region is
  the mirror image; it corresponds to the sequence similar to (c)-(d)
  resulting from an ordering $L_0 <L_1$.}. The top right part of figure \ref{fig-phases} (above $AB_1B_2C$) represents the $2_c$ phase. The region above the line $OD_1B_2A$ (or its mirror image: the region to
the right of the line $OD_2 B_2 C$) corresponds to the $3_c$
phase. The enclosed region $OD_1 D_2O$ corresponds to the $4_c$ phase.
The phase transition across $A B_1$ from right to left corresponds to
the transition $2_c \to 3_c$; the phase transition across  $OD_1$
\footnote{We are treating the non-uniform phase as part of the $4_c$
  phase here.} from above corresponds to $3_c \to 4_c$ (see figure \ref{fig-lattice}).

\begin{figure}[H]
\begin{center}
\includegraphics[scale=.7]{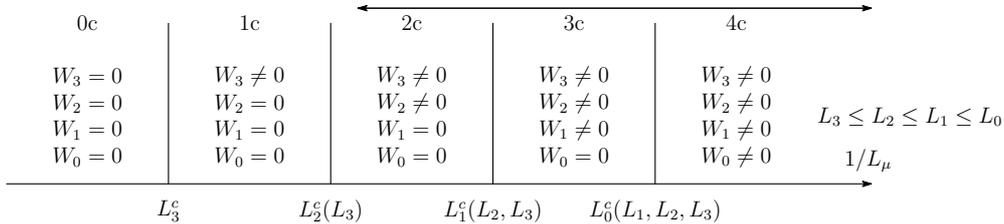}
\caption{Cascade of phase transitions for pure Yang-Mills
theory on a $T^4$ (adapted
from  \cite{Narayanan:2007fb}; see the
points (a)-(e) above for more details). Our results in this paper,
for the theory \eq{d=2} with $D=2$, describe the indicated phases
$2_c, 3_c$ and $4_c$. The $0_c \to 1_c$ transition is found to
be first order from
lattice studies in \cite{Kiskis:2005hf}; the $2_c \to 3_c$
transition 
is found to be first order from our analysis; the $3_c 
\to 4_c$ transition for an asymmetric torus is found in 
our analysis to
be a double (2nd order + 3rd order) transition for
 an appropriate parameter
regime  (see figure \ref{fig-phases}), although, it can be a
single first order transition at other regimes \cite{Mandal:2009vz, Koren2009}.}
\label{fig-lattice}
\end{center}
\end{figure}

The comments (g)-(h) above have a direct bearing on the possible
joining pattern in figure \ref{fig-phases}. Out of the three possible
joining patters shown in the insets, the first and the third patterns
allow a direct transition $2_c \to 4_c$ and are, hence, inconsistent
with the comment (h). Thus, consistency with the lattice results
described in this subsection uniquely pick up the second joining
pattern. As we showed in Section \ref{sec-gravity}, the same joining pattern
is also picked up uniquely in figure \ref{fig-phases-gr} through
the analysis of the gravity dual. 

A more quantitative comparison of our work with the above lattice
studies is left for the future.

We should mention another important lattice study
\cite{Catterall:2010fx} which deals with super Yang Mills theories in
$d=2$ and is closely related to the work presented here and in
\cite{Mandal:2009vz}. In parameter ranges 
where the two theories coincide, our phase diagrams agree
(see section 5 of \cite{Catterall:2010fx}).
See also related numerical works about the center symmetry breaking in super Yang-Mills \cite{Hanada:2009hq, Azeyanagi:2010ne}.

\subsection{\label{other-analytic}Comparison with earlier analytical studies 
with massive adjoint scalars}

Reference \cite{Aharony:2005ew} considers the theory (\ref{d=2}) in
the limit $m \gg {\l_2}^{1/2}$.  Our figure \ref{fig-phases} is
similar to figure 13 of \cite{Aharony:2005ew}, except that our figure
is obtained for any mass (including $m=0$) where their figure is for
the large mass limit.  The reason for the agreement is the appearance
of a dynamical mass $\Delta$ for the adjoint scalars in our model, as
we have explained above.

In figure 14 of \cite{Aharony:2005ew} an interpolation between small
and large radii is proposed on the basis of some analytical estimates
for the intermediate radii.  Our figure \ref{fig-phases-gr}, although
similar to this figure, differs in one crucial respect. The phase
transition line BC in our figure is horizontal throughout, as it must
be according to large $N$ volume independence arguments
\cite{Eguchi:1982nm, Gocksch:1982en}. 
Since the line BC is entirely in
the Tr $V=0$ phase, the transition temperature $\b_c$ cannot depend on
$L$; hence BC must be horizontal.  This property is  violated by the
corresponding line (the intermediate radius segment) of figure 14 in
\cite{Aharony:2005ew}, which should have been horizontal according to the above argument.

\subsection{\label{topology}Comparison with Yang-Mills theories on 
different topologies}

We have found that the nature (in particular, {\em order}) of the
confinement/deconfinement type transition at $\beta_c$ in the two
dimensional gauge theory (\ref{d=2}) at a fixed radius $L$ depends on
the value of $L$. Since this theory can be obtained from a pure
Yang-Mills theory on a small $T^D \times S^1_L\times S^1_\beta$, it is
interesting to compare this result with Yang-Mills theories on compact
spaces with other topologies. We present such a comparison in Table
\ref{tab-topology}\footnote{In order to apply $1/D$ expansion, we need
  $D\ge 1$ in the small $T^D \times$ small $S^1$ case and $D \ge 2$ in
  the small $T^D \times$ large $S^1$ case.}.

\begin{table}[H]
\begin{center}
\begin{tabular}{cc}
\hline $ {\mathcal M} $ & type of phase transition 
\\ \hline 
small $T^D \times$ small $S^1$ & 2nd+3rd 
\\ small $T^D \times$ large $S^1$ & 1st 
\\ small $S^2$ & 2nd+3rd 
\\ small $S^3$ & 1st 
\\ \hline
\end{tabular}
\label{tab-topology}
\caption{Confinement/deconfinement type transitions in pure Yang-Mills
  theories on $S^1_\beta \times {\mathcal M}$.  Here ``small $S^1$''
  and ``small $T^D$'' refer to sizes small enough to ensure (a) that
  the $Z_N$ symmetries in the $S^1$ and $T^D$ directions,
  respectively, are broken, and (b) that all the KK modes can be
  integrated out.  ``Large $S^1$'' ensures that the $Z_N$ symmetry
  along the $S^1$ is not broken.}
\end{center}
\end{table}

Because of the difficulty of the analysis of Yang-Mills theory, only
weakly coupled Yang-Mills theories on $S^2$ and $S^3$
\cite{Aharony:2005bq, Papadodimas:2006jd} have been studied.  In these
cases, all the spatial components of the gauge field have a mass
proportional to $1/R$, where $R$ is the radius of the sphere.  These
massive gauge fields can be integrated out perturbatively if the radius is
sufficiently small ($R \Lambda_{QCD} \ll 1$). Reference
\cite{Aharony:2005bq, Papadodimas:2006jd} derived effective potentials
for $A_0$ up to three loop order and found the transition in the $S^3$
case to be first order \cite{Aharony:2005bq}.  On the other hand, the
transition in the $S^2$ case consists of second and third order
transitions as we found in the small $T^D$ case
\cite{Papadodimas:2006jd}.  Note that the higher order transitions for
small $S^2$ is expected to change to a first order transition in a
strong coupling regime according to lattice studies
\cite{Liddle:2005qb}.

Thus the nature of the transition depends not only on the size but
also on the topology of the compact space.  It would be interesting to
understand the origin of these differences.

\section{\label{Conclusions}Conclusions}

In this work, we have computed the phase diagram of two dimensional
Yang-Mills theory with adjoint scalars (\ref{d=2}), which can be
obtained from a KK reduction of a higher dimensional pure Yang-Mills
theory. We treated the case of massless adjoint scalars in detail,
outlining the generalization to arbitrary non-zero mass in 
Appendix \ref{app-massive}, and found the phase diagram in figure
\ref{fig-phases}. At large spatial radius, there is a first order
confinement/deconfinement phase transition, whereas at small spatial
radius, there are two closely spaced phase transitions: (a) a second
order phase transition from the `confined' phase to a `non-uniform'
phase (non-uniform eigenvalues of the Polyakov line), followed by (b)
a third order phase transition from the `non-uniform' phase to a
`gapped' phase.
 Our calculations, based on the large $D$ method
\cite{Mandal:2009vz}, provide an analytical derivation of the
dependence of the thermodynamic behaviour on the size of the spatial
box, which is anticipated on the basis of lattice studies and
gauge/gravity duality. 

We have also considered the phase transitions in the gauge theory from
the viewpoint of a gravity dual, based on a scaling limit of
Scherk-Schwarz compactification of a D2 brane on a 3-torus. Although
there is no overlapping region of validity of the gauge theory and
gravity descriptions, the analysis of the gravity dual leads us to
conjecture a certain specific completion of the phase diagram in the
gauge theory, as in figure \ref{fig-phases-gr}. In performing this
analysis, we encountered an inherent problem with the holographic
analysis, namely a dependence of the physics
on the fermion boundary conditions, which was absent
in the gauge theory description (see Appendix \ref{app-grav}). 
Indeed this problem is related to a more general problem in the holographic
description of  QCD \cite{Aharony:2006da}.
We discuss this problem further in \cite{Mandal:2011ws}. 

We matched our findings from gauge theory regarding the {\em order} of
various phase transitions with those from a gravity analysis in
Section \ref{sec-gravity} and with those from lattice studies in
Section \ref{other}.

Note that the method of integrating out the adjoint scalars using a
$1/D$ expansion works equally well in higher dimensional ($d\ge 3$)
gauge theories (\ref{d=d}), leading to an effective action for the
gauge field as shown in (\ref{effective-general}).  However it is
difficult to evaluate the dynamics of this model because of the
existence of dynamical gluons.  This is a crucial difference from the
lower dimensional cases ($d=0,1,2$).  In addition, the $d$ dimensional
model (\ref{d=d}) typically appears through a KK reduction of a $d+D$
dimensional (super) Yang Mills theory, but for $d\ge 3$ the mass of
the adjoint scalars induced from loops of KK modes is large (see
appendix \ref{KK loop}); hence the contribution of the adjoint scalars
may be not relevant for $d\ge 3$.

The investigations in the present paper were partly motivated by a
desire to understand a gauge theory dual to a dynamical
Gregory-Laflamme transition. The considerations in this paper provide a
step towards understanding this issue; details of this will appear
elsewhere \cite{progress}.

\subsection*{Acknowledgement}

We would like to thank Gyan Bhanot, Sumit Das, Saumen Datta, Avinash Dhar, Sourendu
Gupta, Shiraz Minwalla, Sreerup Raychaudhuri and Mithat Unsal for
useful discussions. We would also like to thank Pallab Basu,
Manavendra Mahato and Spenta Wadia for numerous interactions and
collaboration at an early stage, and Sumit Das, Avinash Dhar, Shiraz
Minwalla, Rajamani Narayanan, Vasilis Niarchos and Mithat Unsal for
comments on the draft and for pointing out important
references. G.M. would like to thank University of Kentucky, Lexington
for hospitality where part of the work was done, and the seminar
audiences in Lexington, and at the String Theory meeting ISM 2011 in
Puri, for useful interactions.  T.M. would like to thank KEK for
hospitality where part of the work was done.  This work is partially
 supported by Regional Potential program of the E.U. FP7-REGPOT-2008-1:
CreteHEPCosmo-228644 and by Marie Curie contract
PIRG06-GA-2009-256487.

\appendix 

\section{Mass for $Y^I$ from one-loop of the KK modes
\label{KK loop}}

If we consider a $d+D$ dimensional Yang-Mills on $T^{d+D}$ (\ref{d=d+D}) and consider a dimensional reduction by taking the radii of $T^D$ to be small, we will classically obtain a $d$ dimensional gauge theory coupled to $D$ massless adjoint scalars.
However, if we consider quantum effects, the action would be modified.
One of the relevant corrections is that the adjoint scalars would acquire mass as in (\ref{d=d}).
In this appendix, we evaluate the mass at a one-loop level \footnote{We do
not use the large $D$ limit in this appendix.}.

Starting from the $d+D$ dimensional action (\ref{d=d+D}), we can derive a one-loop effective action for the constant diagonal components of $A^\mu = (a^\mu_1,\cdots,a^\mu_N)$ by integrating out all the other modes
\cite{Gross:1980br, Hosotani:1988bm, Aharony:2005ew},
\begin{align} 
S_{eff}=&- \left(\prod_{\mu=0}^{d+D-1}L_\mu \right) \frac{d+D-2}{2}\frac{\Gamma((d+D)/2)}{  \pi^{(d+D)/2}}  \nn
& \times \sum_{i,j}\sum_{\{k_\mu \} \neq \{0\}} \frac{\exp\left(i \sum_\nu k_\nu L_\nu (a^\nu_i-a^\nu_j)\right) }{\left(\sum_\nu k_\nu^2 L_\nu^2 \right)^{(d+D)/2} }. 
\label{effective-zero}
\end{align}
Here the sum $\sum_{\{k_\mu \} \neq \{0\}}$ includes all integers $k_\mu$ except $k_0=\cdots = k_{d+D-1}=0$.
Let us now take $L_\mu $ ($\mu=0,\cdots,d-1$) large and $L_I$ ($I=d,\cdots,d+D-1$) small and derive a low energy effective theory by using this expression.
Gauge invariance implies that the effective action in this situation will be given by
\begin{align} 
S_d = &\int_{0}^\beta \kern-5pt dt \prod_{i=1}^{d-1}
\int_0^{L_i}\kern-5pt dx^i \,  \Tr \left(\frac{1}{4{g_d}^2} F_{\mu\nu}^2
+ \sum_{I=1}^D \frac12 \left(D_\mu Y^{I}\right)^2 -
\sum_{I,J} \frac {{g'_d}^2}{4} [Y^I,Y^J][Y^I, Y^J] \right)\nn
&- \int_{0}^\beta \kern-5pt dt \prod_{i=1}^{d-1}
\int_0^{L_i}\kern-5pt dx^i \,  \left(\prod_{I=1}^D l_I \right) \frac{d+D-2}{2}\frac{\Gamma((d+D)/2)}{  \pi^{(d+D)/2}}  \sum_{\{k_I \} \neq \{0\}} \frac{\left| \Tr e^{i g_d \sum_J k_J l_J Y^J} \right|^2}{\left(\sum_J k_J^2 l_J^2 \right)^{(d+D)/2} }
\label{effective-2}.
\end{align} 
Here we have rewritten $A_{d+I-1}=g_d Y_I$ and $L_{d+I-1}=l_I$.
$g_d$ and $g_d'$ are the same as $g$ and $g'$ of \eq{d=2}; they
satisfy  $g_{d+D}^2/\prod_{I=1}^D l_I=g^2_d=g'^{2}_{d}$ at a physical scale $\mu \gg 1/l_I$ ({\em cf.} \eq{choice}).

If $l_I$ are small and the long string modes are suppressed (see
footnote \ref{tricky}), we can treat $Y^I$ perturbatively.  Then we
can expand the exponentials in (\ref{effective-2}) and obtain a
quadratic term in $Y$ as
\begin{align} 
g_d^2 N \left(\prod_{I=1}^D l_I \right) 
(d+D-2)\frac{\Gamma((d+D)/2)}{  \pi^{(d+D)/2}}
  \sum_{\{k_I \} \neq \{0\}} 
  \frac{ k_I^2 l_I^2  }
  {\left(\sum_J k_J^2 l_J^2 \right)^{(d+D)/2} }
  \frac{\Tr (Y^I)^2}{2}. 
\end{align} 
Other interaction terms from the exponential would be suppressed by
small $l_I$.  If we take all the $l_I$ to have a common value
$l_{KK}=1/M_{KK}$, we obtain a mass for $Y^I$ as
\begin{align} 
m^2=g^2_dN M^{d-2}_{KK} (d+D-2)\frac{\Gamma((d+D)/2)}{ \pi^{(d+D)/2}}
\sum_{\{k_I \} \neq \{0\}} \frac{1}{D} \frac{ 1 }{\left( k_1^2+\cdots
  +k_D^2 \right)^{(d+D)/2-1} }.
\label{KKmass}
 \end{align} 
This sum would diverge for $d\le 2$ and hence needs to be regulated 
e.g.  by using a prescription $(\vec k)^2
\equiv k_1^2+\cdots +k_D^2 \le (\Lambda_{UV}/
M_{KK})^2 $, where $\L_{UV}$ is a cut-off scale \footnote{An ultraviolet
cutoff typically breaks gauge symmetry. The calculation discussed here
can be repeated avoiding such problems, by using dimensional regularization \cite{Luscher:1982ma}. }.
 This would imply a non-trivial RG flow of the mass.  Let us first
consider the case of $d=2$. For $\L_{UV} \gg M_{KK}$, the regulated
sum \eq{KKmass} approximately gives $ m^2(\L_{uv}) = A~ g^2_2 N 
\log(\L_{uv}/M_{KK}) $ where $A$ is a numerical constant.  Let us
choose a renormalization scale $\mu \gtrsim O(M_{KK})$. We can then
define a renormalized mass at the scale $\mu$ as 
\begin{align}
{\bar m}^2 
\equiv  m^2(\L_{uv}) - A~ g_2^2 N  \log(\L_{uv}/\mu) = A' \lambda_2
\label{m-ren},
\end{align}
where $A'= A \log(\mu/M_{KK}), \l_2 = g_2^2 N$.  
Note that the running of the mass will stop below $\mu < M_{KK}$.

For $d=1$, a similar analysis gives 
\begin{align}
{\bar m}^2 \sim \lambda_3  /M_{KK}
\label{m-ren-1}.
\end{align}

For $d\ge 3$, the sum in \eq{KKmass} is convergent, leading to
$m \sim \lambda_d^{1/2} M_{KK}^{(d-2)/2}$  which is  much larger than the typical QCD scale, e.g. for $d=3$ the QCD scale is $O(\lambda_3)$ whereas the adjoint mass is $O(\lambda_3^{1/2} M_{KK}^{1/2})$.
Thus if $M_{KK}$ is large, the adjoint scalars will not contribute to the QCD dynamics.
It means that only the $d$ dimensional gauge field dominates the dynamics\footnote{For $d=2$, the mass of the adjoint scalar from the KK modes is finite
($O(\l_2)$) but the dynamical mass $\Delta$ is (logarithmically) larger than
$\l_2$ (see \eq{choice2}).
Hence one may naively think that the adjoint scalar would be irrelevant as for $d\ge3$.
However this is not correct \cite{Aharony:2005ew}, since the phase structure of the 2d pure Yang-Mills on $T^2$ is trivial and is always confined. Therefore the contribution of the (logarithmically) heavy adjoint scalars is important in the 2 dimensional gauge theory (\ref{d=2}). 
}.

So far we have considered the mass correction from the KK modes in pure Yang-Mills theory.
We can extend this calculation to the KK reduction of supersymmetric theories with a Scherk-Schwarz compactification e.g. to the derivation of (\ref{d=2}) with $D=8$ from the D2 theory.
In this case, we need to evaluate the contribution of loops of adjoint scalars as well as fermions.
However, it can be shown that we obtain a similar mass $\sim O(\lambda_d^{1/2} M_{KK}^{(d-2)/2})$ even in this case \cite{Aharony:2005ew}.

\section{\label{app-derivation} 
Derivation of effective potential \eq{arbit-u-v}}

In this appendix, we show the derivation of the effective action
(\ref{arbit-u-v}) from (\ref{logdet}) with the assumptions (a) and (b)
below \eq{logdet}.  We can evaluate (\ref{logdet}) in a general $d$
dimensional gauge theory, described by the action
\begin{align} 
\kern-5pt S = \int_{0}^\beta \kern-5pt dt \prod_{i=1}^{d-1}
\int_0^{L_i}\kern-5pt dx^i \, \Tr \Biggl(&\frac{1}{4{g}^2} F_{\mu\nu}^2 
+ \sum_{I=1}^D \frac12 \left(D_\mu Y^{I}\right)^2 
+\frac{m^2}{2}Y^{I2}  \nn &
-
\sum_{I,J} \frac {{g'}^2}{4} [Y^I,Y^J][Y^I, Y^J] \Biggr).
\label{d=d}
\end{align}
This model can be identified with (\ref{effective-2}) if we take $L_I$
small and choose the mass and couplings appropriately.  We will first
discuss the general $d$-dimensional case and apply the results to
$d=2$ later.

Let us first set $m=0$.  Then, through a similar calculation as in
Section \ref{sec-large-D}, we obtain a generalization of
(\ref{logdet}) (the calculation closely follows \cite{Aharony:2005ew}).
In this section, we will use the more general notation $(L_0, L_1)$
for $(\beta,L)$.
\begin{align}
\delta S[A, \Delta]=&\frac{D}{2} \log \det \left( -D^2_\mu +\Delta^2 \right) =
\frac{D}{2} \Tr \sum_{\{n_\mu\}} \log \left( \sum_{\mu=0}^{d-1} \left(\frac{2\pi n_\mu}{L_\mu}+A_\mu
\right)^2 +\Delta^2 \right)
\nonumber \\ =&
 \frac{D}{2} \frac{L_0\cdots L_{d-1}}{(2\pi)^d} 
 \Tr_{adj} \sum_{\{k_\mu\}}P_{\{k_\mu\}}(\Delta,\{L_\mu\}) e^{i \sum_\mu k_\mu L_\mu A_\mu},
 \label{logdet-cal}
\end{align} 
where $P_{\{k_\mu\}}$ is 
\begin{align} 
&P_{\{k_\mu\}}(\Delta,\{L_\mu\}) \nonumber \\ 
=&
\int_0^{2\pi/L_0 }da_0 \cdots
\int_0^{2\pi/L_{d-1} }da_{d-1} 
 \sum_{\{n_\mu\}} \log \left(
 \sum_{\mu=0}^{d-1} \left(\frac{2\pi
    n_\mu}{L_\mu}+a_\mu \right)^2
  +\Delta^2 \right)e^{-i \sum_\mu k_\mu L_\mu a_\mu}\nn
  =&
\sum_{\{n_\mu\}}
\int_{n_0}^{n_0+2\pi/L_0 }da_0 \cdots
\int_{n_{d-1}}^{n_{d-1}+2\pi/L_{d-1} }da_{d-1} 
  \log \left(
 \sum_{\mu=0}^{d-1} a_\mu^2
  +\Delta^2 \right)e^{-i \sum_\mu k_\mu L_\mu a_\mu}\nn
   =&
\int_{-\infty}^{\infty }da_0 \cdots
\int_{-\infty}^{\infty }da_{d-1} 
  \log \left(
 \sum_{\mu=0}^{d-1} a_\mu^2
  +\Delta^2 \right)e^{-i \sum_\mu k_\mu L_\mu a_\mu}.
\end{align} 
Let us now evaluate $P_{\{k_\mu\}}$ in the $\{k_\mu\}=\{0\}$ and
$\{k_\mu\}\ne \{0\}$ cases separately.
In the $\{k_\mu\} \neq \{0\}$ case, $P_{\{k_\mu\}}(\Delta,\{L_\mu\})$ becomes, 
\begin{align} 
&P_{\{k_\mu\}}(\Delta,\{L_\mu\}) \nn
  =& 
\int_{-\infty}^{\infty }da_0 \cdots
  \int_{-\infty }^{\infty }da_{d-1}  \log \left( \sum_{\mu=0}^{d-1} a_\mu^2 +\Delta^2 \right)e^{-i \sum_\mu k_\mu L_\mu a_\mu}  \nn
=& 
\int_{-\infty}^{\infty }da_0 \cdots
  \int_{-\infty }^{\infty }da_{d-1} 
 \lim_{\epsilon \to 0}  \left[-\log \epsilon -\gamma  - \int_\epsilon^\infty \frac{d\alpha}{\alpha} 
   e^{-\left( \sum_\mu a_\mu^2 +\Delta^2 \right)\alpha}
   \right]
   e^{-i \sum_\mu k_\mu L_\mu a_\mu}  \nn  
=& 
   - \pi^{d/2} \int_0^\infty \frac{d\alpha}{\alpha^{d/2+1}} 
   e^{-\Delta^2 \alpha- \frac{1}{4\alpha}  \sum_\mu( k_\mu L_\mu)^2 }  \nn     
 =&-\frac{2 \left(2\pi \Delta \right)^{d/2} }{ \left( \sqrt{\sum_\mu \left( L_\mu k_\mu\right)^2}  \right)^{d/2}  } K_{\frac{d}{2} }
 \left(\Delta \sqrt{\sum_\mu \left( L_\mu k_\mu\right)^2} \right) \nn
 =&-\frac{2 \left(2\pi \Delta \right)^{d/2} }{ \left( \sqrt{\sum_\mu \left( L_\mu k_\mu\right)^2}  \right)^{\frac{d+1}{2} }  }
\sqrt{\frac{\pi}{2\Delta} }
 \exp
 \left(-\Delta  \sqrt{\sum_\mu \left( L_\mu k_\mu \right)^2} \right)  +\cdots,
 \label{P_k}
  \end{align} 
where $K_{d/2}$ is the modified Bessel function of the second kind and
we have used $K_{a}(z)=\sqrt{\pi/2z}e^{-z}+\cdots$ for  large $z$ in
the last equation.

For $\{k_\mu\}=\{0\}$, we find
\begin{align} 
P_{\{0\}}(\Delta,\{L_\mu\})=& \int_{-\infty}^{\infty }da_0 \cdots
  \int_{-\infty }^{\infty }da_{d-1}  \log \left( \sum_{\mu=0}^{d-1} a_\mu^2 +\Delta^2\right) \nn
  =&  \int_{0}^{\Lambda }da \Omega_{d}  a^{d-1}  \log \left(  a^2 +\Delta^2  \right),
  \label{P_0}
\end{align} 
where $\Omega_{d}=2\pi^{d/2}/\Gamma(d/2)$ is the surface area of the
$d$ dimensional unit sphere and $\Lambda$ is a cut off.

Finally we turn on the mass term.  In this case, we can obtain the
results by replacing $\Delta \to \sqrt{\Delta^2+m^2}$ in (\ref{P_k})
and (\ref{P_0}).  Note that the assumption (a) and (b) below
(\ref{logdet}) should be modified accordingly.

\subsection{Effective action for $d=2$}

We now consider the special case of $d=2$.  
In this case, we can evaluate $P_{\{0\}}$ as 
\begin{align} 
P_{\{0\}}(\Delta,\beta,L)=& 2\pi \int_{0}^{\Lambda }da~ a \log \left(
a^2 +\Delta^2 \right) \nn
 =&-\pi\Lambda^2 +\pi \left(  \Lambda^2+\Delta^2 \right)  \log
\left( \Lambda^2+\Delta^2 \right) 
-\pi \Delta^2 \log \Delta^2 .
\end{align} 
By using this result and (\ref{P_k}) to (\ref{logdet-cal}), we
obtain the effective action (\ref{arbit-u-v}).  Note that $\Tr_{adj}
e^{i( k \beta A_0 + l L A_1)} $ in (\ref{logdet-cal}) becomes $\left|
\Tr \left( U^k V^l\right) \right|^2$ by using the assumption (b).

\subsection{Effective potential for higher dimensional models}
\label{app-effective-HD}

It is easy to generalize the derivation of the effective potential
(\ref{Seff-V}) for the two dimensional gauge theory in
Section \ref{sec-large-D} to the $d$ dimensional gauge theory (\ref{d=d})
for large $L_\mu$ \footnote{The $1/D$ expansion in the $d$
  dimensional gauge theory (\ref{d=d}) in a high temperature region is
  considered in \cite{Morita:2010vi} also.}.  By using the results in
the previous section, we obtain
\begin{align} 
S=&
 \prod_{\mu=0}^{d-1}
\int_0^{L_\mu}dx_\mu \left( 
 \frac{1}{4g^2}  \Tr F_{\mu\nu}^2
 -\frac{DN^2}{(2\pi)^{\frac{d-1}{2}}}
\sum_{\mu=0}^{d-1} \frac{(\Delta_0^2+m^2)^{\frac{d-1}{4}  }}{L_\mu^{\frac{d+1}{2} }}  e^{-\sqrt{\Delta_0^2+m^2} L_\mu} \left| \frac{1}{N} \Tr e^{iL_\mu A_\mu} \right|^2  \right) \nn
&+DN^2L_0 \cdots L_{d-1}\left(-\frac{\Delta_0^4}{8\tilde{\lambda}' }  
+\frac{1}{2(2\pi)^d}P_{0}\left(\sqrt{\Delta_0^2+m^2},\{L_\mu\} \right)  
\right) .
\label{effective-general}
\end{align} 
Here $\Delta_0$ is defined as a solution of the saddle point equation
\begin{align} 
\frac{\Delta^2}{2\tilde{\lambda}' } = \frac{\Omega_d}{(2\pi)^d} \int_0^\Lambda da \frac{a^{d-1}}{a^2+\Delta^2+m^2} .
\end{align} 
Note that we can always find a unique positive solution $\Delta_0$
from this equation.

\section{The phase structure of massive model.}
\label{app-massive}

In this appendix, we study the two dimensional gauge theory
(\ref{d=2}) with a mass term for the adjoint scalars. As we mentioned
in the introduction and elaborated in appendix \ref{KK loop},
such a mass term generically arises from KK loops.
We discuss here how the results of the massless case in
Section \ref{sec-large-L} and Section \ref{sec-small-L} are modified.  We
will show that the mass does not change the qualitative nature of the
phase structure.

\subsection{Large $L$ case}

By using the results in \ref{app-effective-HD}, we generalize the
effective action for $\Tr U$ (\ref{eff-action}) to the massive case.
The resulting effective action is again given by 
(\ref{eff-action}) with different values of $\xi$ and $C$:
\begin{align} 
\xi=& \sqrt{\frac{\sqrt{\Delta_0^2+m^2}}{2\pi \tilde{\lambda}^2
    \beta^3}}e^{- \beta \sqrt{\Delta_0^2+m^2}} ,\\
C(\tilde \l', \Delta_0)=& \frac{\beta L \Delta_0^2}{8\pi}\left(1+
\frac{\pi \Delta_0^2}{\tilde{\lambda'} }   \right) 
+\frac{\beta L m^2}{8\pi}\left(1+
\frac{2\pi \Delta_0^2}{\tilde{\lambda'} }   \right). 
\end{align}
The dynamical mass $\Delta_0$ is a solution of
\begin{align} 
\tilde{\lambda'}
=\frac{2\pi \Delta_0^2}{\log\left(1+\frac{\Lambda^2}{\Delta_0^2+m^2}   \right)  }.
\label{saddle-massive}
\end{align} 
For large $\Lambda$, $\Delta_0$ becomes
\begin{align} 
\Delta_0=
\sqrt{\frac{\tilde{\lambda'}}{2\pi} 
W\left(\frac{2\pi \Lambda^2}{
\tilde{\lambda'} } e^{\frac{2\pi m^2}{\tilde{\lambda'}} } \right) -m^2 }.
\end{align} 
Therefore we can use the same analysis as in Section \ref{sec-phase
  tr} and obtain the same phase structure with the following modified
transition temperatures
\begin{align} 
\beta_m=&\frac{3}{2\sqrt{\Delta_0^2+m^2}}W\left[ \frac{2}{3(2\pi \xi_m^2)^{1/3}}
  \left(\frac{\Delta_0^2+m^2}{\tilde{\lambda}} \right)^{2/3} \right].
\end{align}
Although the mass changes the explicit values of the transition
temperatures and some other physical quantities, the qualitative
nature of the phase structure is not modified, as we have mentioned.

Note that (\ref{saddle-massive}) implies that $\Delta_{0}$ becomes smaller as $m$ increases for fixed $\tilde{\lambda}'$ and $\Lambda$.  Therefore, for heavy mass, we can ignore the dynamical mass $\Delta_0$ compared to $m$ and our calculations reproduce the heavy mass QCD results in
\cite{Semenoff:1996xg, Aharony:2005ew}.

\subsection{Small $L$ case}

If $L$ is small enough in (\ref{d=2}), we can integrate out all the
non-zero momentum modes in the $L$ direction (see footnote
\ref{tricky}) and obtain a one dimensional model (\ref{d=d}) with
$d=1$.  In this case, the mass of the adjoint scalars induced at one
loop from the KK modes is proportional to
$(\tilde{\lambda}_{1}L)^{1/2}$ (see (\ref{m-ren-1})). Hence we can ignore it for small enough $L$ and the results in
\cite{Mandal:2009vz} shown in Section \ref{sec-small-L} would still be
valid.  Although the contribution from the mass would be small, it may
be valuable to confirm that the results in \cite{Mandal:2009vz} are
not modified qualitatively.

Starting from (\ref{d=d}) with $d=1$, we obtain an effective potential
through a similar calculation as in
\ref{app-effective-HD}\footnote{\label{ftnt-d1-strong}Contrary to the $d\ge 2$ case, the
  condition (a) and (b) below (\ref{logdet}) are not required for a
  derivation of the effective action in the $d=1$ case.  Thus the
  $1/D$ analysis would be valid as long as the effective dimensionless
  coupling $\tilde{\lambda}_1 \beta^3$ does not scale with $D$.},
\begin{align} 
S_{eff}(\Delta,\{u_n\})/DN^2 
 =- \frac{\beta \Delta^4}{8 \tilde{\lambda}_1  } 
+ \frac{\beta \sqrt{\Delta^2+m^2}}{2} +\sum_{n=1}^\infty
\left(\frac{1}{D}-e^{-n\beta \sqrt{\Delta^2+m^2}} \right) \frac{|u_n|^2}{n} .
\label{d=1-massive}
\end{align} 
Here the third term is a contribution of the Vandermonde determinant \cite{Mandal:2009vz}.
Then the saddle point equation for $\Delta^2$ becomes
\begin{align} 
\frac{\Delta^2}{\tilde{\lambda}_1 }&= \frac{1}{\sqrt{\Delta^2+m^2}} +
2\sum_{n=1}^\infty \left( \frac{1}{\sqrt{\Delta^2+m^2}} e^{-n\beta
  \sqrt{\Delta^2+m^2}} \right) |u_n|^2.
\end{align} 
The solution of this equation at low temperatures ($e^{-\beta
  \sqrt{\Delta^2+m^2}} \ll1$) is
\begin{align} 
\Delta^2=& \Delta^2_0 +\frac{4 \tilde{\lambda}_1 \sqrt{\Delta_0^2+m^2} }{3 
\Delta_0^2+ 2m^2} e^{-\beta \sqrt{\Delta^2_0+m^2}}|u_1|^2+\cdots  ,
\end{align} 
where
\begin{align} 
\Delta^2_0=& \frac{m^2}{3}\left(f(m)+f(m)^{-1} -1 \right), \nonumber \\
f(m)=&\frac{1}{2^{1/3}m^2} \left(
27\tilde{\lambda}_1^2 -2m^6+\sqrt{27\tilde{\lambda}^2_1 (27\tilde{\lambda}^2_1 -4m^6)} 
\right)^{1/3} .
\end{align} 
Note that, although $f(m)$ is a complex for $m^6 \ge
27\tilde{\lambda}_1^2/4$, $\Delta^2_0$ is always real and positive.
By substituting this solution to (\ref{d=1-massive}), we obtain an
effective action for $u_n$,
\begin{align} 
S_{eff}(\{u_n\})/DN^2 
 =- \frac{\beta \Delta_0^4}{8 \tilde{\lambda}_1  } 
+ \frac{\beta \sqrt{\Delta^2_0+m^2}}{2} +
a|u_1|^2+b|u_1|^4 + \cdots,
\end{align} 
where
\begin{align} 
a=\left(\frac{1}{D}-e^{-\beta \sqrt{\Delta^2_0+m^2}} \right), \quad
b=\frac{ \beta \tilde{\lambda}_1 }{3 \Delta_0^2+ 2m^2} e^{-2\beta
  \sqrt{\Delta^2_0+m^2}}.
\end{align} 
One important fact is that $b$ is always positive.  It has been shown
that, in this case, the confinement/deconfinement type transition
always consists of the two transitions (2nd+3rd) as in the massless
case \cite{Mandal:2009vz, Aharony:2003sx}.  These critical
temperatures are given by the solutions of $a=0$ and $a/(2b)=-1/4$
respectively, and evaluated as
\begin{align} 
\beta_{c1}=& \frac{1}{\sqrt{\Delta_0^2+m^2}} \log D,  \nn
\beta_{c2}=&\beta_{c1}- \frac{\tilde{\lambda}_1 }{2D}\frac{\log D}{3 \Delta_0^2+2m^2}  .
\end{align} 
Therefore as in the large $L$ case, the qualitative nature of the
phase transition is not modified by the mass
term\footnote{\label{disagree} Note that our results based on the
  $1/D$ expansion disagree with the results in
  \cite{Aharony:2005ew}. Reference \cite{Aharony:2005ew} studied the same model
  \eq{d=1-massive} by using a large mass approximation, which is
  supposed to be valid if $\tilde{\lambda}_1/m^3 \ll 1 $, up to
  three-loop order and concluded that the confinement/deconfinement
  transition in this model would be a single first order transition.
  The difference would presumably arise from the fact that the $1/D$
  expansion employed here evaluates the model in a non-trivial vacuum
  characterized by the non-zero $\Delta$, whereas the large mass
  analysis of \cite{Aharony:2005ew} is performed as a perturbation
  around the trivial vacuum. Numerical studies analogous to
  \cite{Kawahara:2007fn} but performed for massive adjoint scalars \cite{Azuma:2007fj} should be able to provide further insight into this issue.}.

Although we have evaluated only the leading order of the $1/D$
expansion in this section, the result does not change even if we
include the next order.

\section{\label{app-grav} Phase structure of D2 branes on a 3-torus}

In this appendix, we consider the gravitational system of
Section \ref{sec-gravity} in detail. We first review some generalities
for D$p$ branes with arbitrary $p$.

\subsection{D$p$ branes wrapped on $p+1$-Torus}
\subsubsection{The solutions}
\label{sec-Dp-sol}

The geometry of a black D$p$ brane on a $p$-torus $x_i \in (0,L_i)$,
($i=1,\cdots, p$) in the Maldacena limit (assuming Euclidean time
$t\in (0,\b)$ ) \cite{Itzhaki:1998dd} is given by
\begin{align}
ds^2 &= \a' \left[F(u)
\left(f(u) dt^2 + \sum_{i=1}^p dx_i dx_i)\right)+ \frac{du^2}{F(u)f(u)} 
+ G(u) d\Omega_{8-p}^2\right],
\nn
e^\phi&= \frac{(2\pi)^{2-p}\l_{p+1}}{N}[F(u)]^{(p-3)/2},
\nn
F(u) &= \frac{u^{(7-p)/2}}{\sqrt{d_p\l_{p+1}}}, \quad
G(u)= \sqrt{d_p\l_{p+1}} u^{(p-3)/2}, \quad
f(u)= 1- \left( \frac{u_0}{u} \right)^{7-p},
\nn
d_p&= 2^{7-2p}\pi^{(9-3p)/2}\Gamma\left( \frac{7-p}2 \right) , \quad
\l_{p+1}= g_{p+1}^2 N,
\label{dp-metric}
\end{align}
where $g_{p+1}^2$ is the $p+1$ dimensional YM coupling, which, in the
bulk theory, can be regarded as specifying a boundary condition for
the dilaton field. 
The dimensionless YM coupling, defined by
\begin{align}
g_{eff}^2 = (g_{p+1}^2N) u^{p-3}= (e^\phi N)^{2/(7-p)},
\label{g-eff}
\end{align}
is given directly in terms of the dilaton; its dependence on $u$
for $p\ne 3$ reflects the running of the gauge coupling. 
The scalar curvature is given by
\begin{align}
\a' R = 1/g_{eff}.
\label{ricci-p}
\end{align}
Since time is Euclidean, $u \in (u_0, \infty)$. The smoothness
condition at $u=u_0$ relates the inverse temperature $\b$ to
$u_0$ as follows:
\begin{align}
\frac{\b}{2\pi}= \frac{\sqrt{d_p \l_{p+1}}}{7-p} u_0^{(p-5)/2}
\label{u0-beta}.
\end{align} 

The classical action of the black D$p$ brane is \cite{Itzhaki:1998dd,
  Aharony:2004ig, Hanada:2007wn}
\begin{align} 
S/N^2=& C_p \lambda_{p+1}^{\frac{p-3}{5-p} }L_1 
\cdots L_p \beta \left(- 
\beta^{-\frac{2(7-p)}{5-p}} + H_{\rm reg}(U) \right), 
\nn
C_p&=\frac{5-p}{2^{11-2p}\pi^{(13-3p)/2}\Gamma((9-p)/2) a_p^{2(7-p)/(5-p)}} ,
\nn
a_p&=\frac{7-p}{4\pi d_p^{1/2}}, \quad 
d_p=2^{7-2p}\pi^{(9-3p)/2}\Gamma\left(\frac{7-p}{2}  \right) , 
\nn
H_{\rm reg}(U)& =  \left(\frac{2 a_p}{\sqrt{\l_{p+1}}}\right)^{2(7-p)/(5-p)}
U^{7-p},
\label{black-action}
\end{align}
which is evaluated by expressing the on-shell action as
a regularized integral of $\sqrt{g}e^{-2\phi}R$ over the range
$u_0 \le  u \le U$.

In the following we will also be interested in AdS solitons
\cite{Witten:1998zw, Horowitz:1998ha} which can be obtained from
\eq{dp-metric} by moving the coefficient $f(u)$ from $dt^2$ to one of
the $x_i$'s , say to $dx_p^2$:
\begin{align}
ds^2 = \a' \left[F(u)
\left(f(u) dx_p^2 + dt^2 + \sum_{i=1}^{p-1} 
dx_i dx_i)\right)+ \frac{du^2}{F(u)f(u)} 
+ G(u) d\Omega_{8-p}^2\right].
\label{soliton-metric}
\end{align}
The smoothness condition at $u=u_0$ now gives a condition analogous to
\eq{u0-beta} where $\beta$ is replaced by $L_p$.  Thus this solution
has a contractible $x_p$-cycle (which wraps around  a so-called
`cigar' geometry on the $(x_p, u)$ plane) along which the fermions
must obey the anti-periodic (AP) boundary condition.

The regularized classical action evaluated on such a classical
configuration is given by
\begin{align} 
S/N^2=& C_p \lambda_{p+1}^{\frac{p-3}{5-p} }L_1 \cdots L_p
\beta \left(-L_p^{-\frac{2(7-p)}{5-p}} + H_{\rm reg}(U) \right),
\label{soliton-action}
\end{align}
where the notations are the same as before.

In a toroidal, Euclidean, spacetime, the time direction is on a
similar footing as any other direction.  Thus, the difference between
the black brane and the solitonic solution is only in the labelling of
the contractible cycle (location of the `cigar' geometry).  Hence we
will sometimes refer to both black branes as well as AdS solitons as
just {\it D$p$ solutions}. In the following sections, we will consider
different D$p$ solutions (with $p=0,1,2$) which wrap on (are localized
along) various cycles and, in order to distinguish them, we will use
the following notation: 

$Dp_{L_0(L_1,...,L_p)}$ denotes a D$p$ solution which (i) has a
contractible $L_0$ cycle (that winds around the `cigar') and (ii)
wraps on the $L_1,...,L_p$ cycles; this is a black brane. Similarly
$Dp_{L_p(L_0,L_1...,L_{p-1})}$ denotes an AdS soliton in which the
roles of $t$ and $x^p$ are flipped, as in \eq{soliton-metric}.

\subsubsection{Validity of supergravity}
\label{sec-gr-valid}

The solutions described in the previous section are leading order
supergravity solutions.  When we consider a black D$p$ brane solution
(\ref{dp-metric}), the gravity analysis is reliable, if the parameters
satisfy the following conditions \cite{Harmark:2007md}:
\begin{itemize}

\item[1.] The typical length scale of the black D$p$ brane near the
  horizon (see, e.g. \eq{ricci-p}) is given by $l= \left(\alpha'
  \sqrt{d_p \lambda_{p+1}} u_0^{(p-3)/2}\right)^{1/2}$.
  In order to suppress stringy excitations, we should satisfy $l \gg
  \sqrt{\alpha'}$. 
  From (\ref{u0-beta}), this condition is equivalent to 
    \begin{align} 
     \lambda_{p+1} \beta^{3-p} \gg 1 \quad (p \le 5).
     \label{cond-string}
    \end{align} 

\item[2.] The mass of the winding mode along an $L_i$ cycle is given
  by $M_{wi}=\left(\alpha' u_0^{(7-p)/2}/\sqrt{d_p
    \lambda_{p+1}}\right)^{1/2}L_i/\alpha'$.  
    In order to suppress the winding mode, we must have $M_{wi} l \gg 1$.  This condition gives
  \begin{align} 
    \lambda_{p+1}^{1/2}L_i^{\frac{5-p}{2} } \gg \beta \quad (p \le 5).
    \label{cond-winding}
    \end{align}   
    If this condition is violated and if the fermion on the brane satisfies the periodic (P) boundary condition along the $L_i$ cycle, we can perform a $T$-duality along this direction and reassess the validity of supergravity in the dual frame\footnote{If the fermion satisfies an anti-periodic (AP) boundary condition along the $L_i$ cycle, the theory is mapped to a type 0 theory through the $T$-duality. Then the bulk theory involves a tachyon and how the holographic description of gauge theory works in this frame is unclear. We thank Shiraz Minwalla for pointing this out.}.
      After the $T$-duality, the black D$p$ solution
    becomes a smeared black D$(p-1)$ brane solution, which is composed
    of uniformly distributed D$(p-1)$ branes on the dual $L_i$ cycle \cite{Aharony:2004ig, Harmark:2004ws, Harmark:2005jk}.
    Then the condition (\ref{cond-winding}) is replaced by
    $\tilde{M}_{wi} l \gg 1$, where 
    $\tilde{M}_{wi} \equiv \left(\alpha'
    u_0^{(7-p)/2}/\sqrt{d_p \lambda_{p+1}}\right)^{-1/2}(2\pi)^2/L_i$ 
     is the mass of the winding mode on the dual cycle.  This
    condition gives
  \begin{align}
  L_i \ll \beta. 
  \label{cond-dual}
  \end{align} 
  Note that the first condition (\ref{cond-string}) does not change under the $T$-duality
(the value of the classical action is also invariant).
\end{itemize}

If, instead of a black D$p$ brane, we consider a solitonic solution
which is obtained from the black brane by flipping $t \leftrightarrow
x_i$, then the conditions for the validity of supergravity are simply
obtained by replacing $\beta$ and $L_i$ in the above conditions.

We will discuss these criteria below in some detail in the parameter
regime of interest.

\subsection{Phase transitions of the D$p$ solutions}
\label{sec-transition-general}

Using the on-shell classical actions (\ref{black-action}) and
(\ref{soliton-action}), we can determine various phase transitions
\cite{Martinec:1998ja, Aharony:1999ti, Hanada:2007wn}, as we will show
now.

\subsubsection{GL transition}

The Gregory-Laflamme (GL) transition \cite{Gregory:1994bj, Kol:2004ww, Harmark:2007md} was
originally found in the context of black rings in $D\ge 5$ which were
found to be unstable unless wrapped on a sufficiently small circle.
In the context of black $p$-branes, the GL instability shows up as
follows \cite{Aharony:2004ig, Harmark:2004ws, Harmark:2005jk}.
Suppose a black $p$-brane is wrapped on a circle of length $L_1$, along which the fermion satisfies the periodic boundary condition.
If $L_1$ is so small that it violates (\ref{cond-winding}), we need the $T$-dualized
description in terms of a uniformly smeared black $(p-1)$-brane on the
dual circle of length  $L'_1= (2\pi)^2/L_1$.  If $L'_1$ is large enough, the
smeared black $(p-1)$-brane undergoes a GL transition leading to a black $p-1$
brane localized on the dual cycle.  The transition can be studied
dynamically, as well as thermodynamically. To study the latter, let us
interpret \eq{black-action}  without the regulator term  (see Eq. (9) of
\cite{ Itzhaki:1998dd}) as the 
Euclidean action (above extremality) of the smeared
black $(p-1)$-brane :
\begin{align}
S_p /N^2= - C_p \lambda_{p+1}^{\frac{p-3}{5-p} }L_1 
\cdots L_p \beta \beta^{-\frac{2(7-p)}{5-p}}.
\label{black-nonextremal}
\end{align}
Here the action is evaluated in the D$p$-frame (recall that the action
is invariant under $T$-duality).
 
The value of the Euclidean action (above extremality) for the
localized black $p-1$ brane solution is approximately (see below)
given by\footnote{Improvements to this approximation 
are discussed in \cite{Harmark:2004ws}.}
\begin{align}
S_{p-1} /N^2= - C_{p-1} \lambda_{p}^{\frac{p-4}{6-p} }L_2 \cdots
L_{p}  \beta \beta^{-\frac{2(8-p)}{6-p}}.
\label{black-p-1}
\end{align}
For small enough $L_1$ (large enough $L'_1$), this is smaller than
\eq{black-nonextremal}.  The transition between the uniform and
localized $p-1$ branes happens when \eq{black-nonextremal} equals
\eq{black-p-1}:
\begin{align}
\frac{\beta}{L_1}=  \left(\frac{C_p}{C_{p-1}}
\right)^{(5-p)(6-p)/4} \sqrt{ \l_{p+1} L_1^{3-p}}
\label{gl-value}.
\end{align}
Here we have used $\l_p=\l_{p+1}/L_1$.  In \eq{black-p-1} one has 
used the approximation that the horizon size is much smaller than
$1/L_p$. Since this is strictly not true near the phase transition
point (where the horizon size is of the order of $1/L_1$), the
estimate \eq{gl-value} is approximate.
See \cite{Hanada:2007wn} for some details of this approximation.

Note that there are several arguments that this transition would be
first order \cite{Aharony:2004ig, Harmark:2007md, Hanada:2007wn}.

\subsubsection{GL type transition in the soliton sector}
\label{app-GLtype}

For bosonic theories in Euclidean spacetimes, the $\b \leftrightarrow
L_p$ interchange is a symmetry, provided all other radii are left
unchanged.  This is a symmetry of fermionic theories as well, provided
the spin structures along $t$ and $x^p$ respect this symmetry. Thus,
if there is a GL transition given by \eq{gl-value}, there must be a GL
type transition between a uniformly distributed solitonic $(p-1)$ brane
to a localized solitonic $(p-1)$ brane when $1/L_1$ becomes too
large. The transition is given by \eq{gl-value} with a $\b
\leftrightarrow L_p$ interchange:
\begin{align}
\frac{L_p}{L_1}=  \left(\frac{C_p}{C_{p-1}}
\right)^{(5-p)(6-p)/4} \sqrt{\l_{p+1} L_1^{3-p}}
\label{gl-s-value}.
\end{align}

\subsubsection{The Scherk-Schwarz (SS) transition}

This is a transition between the black $p$ brane configuration
\eq{dp-metric} and its solitonic counterpart \eq{soliton-metric}. 
If we use the same large $U$ regulator for both the solutions, then
\eq{black-action} and \eq{soliton-action} can easily be compared. The
regulator term is the same in both the actions and can be ignored
while comparing the two. Equating the two Euclidean actions, we
get the transition temperature
\begin{align}
\b =  L_p.
\label{ss-value}
\end{align}
Similarly, we can consider another solitonic solution by replacing
$L_p \leftrightarrow L_k$, if $L_k$ is also an AP circle.  Then
further transitions will happen at $\beta=L_k$ and $L_p=L_k$.

Note that this transition is also expected to be first order \cite{Aharony:1999ti}.

\subsection{D2 branes for various spin structures: generalities}

In this section, we apply the general properties of the D$p$ solutions
studied in the previous sections to the D2 brane on the 3-torus:
$(t,x_1,x_2)=(t+\beta,x_1+L_1,x_2+L_2)$ \footnote{The notation $L_1$ in this section represents what is called $L$ in the main text.} and make a prediction for the
phase structure of the 2 dimensional gauge theory (\ref{d=2}).  As
mentioned in Section \ref{sec-gravity}, we fix AP boundary condition
for fermions along the $x_2$ circle.  Then, there are 4 choices of 
boundary conditions
for the fermions along the $t$ and $x_1$ directions: (AP,AP),
(AP,P), (P,AP), (P,P), where P denotes the periodic boundary condition
and AP denotes the anti-periodic one.  We will evaluate the phase
structure of the gravitational system with these 4 boundary
conditions. (Since the result in the (P,AP) case can be obtained from
(AP,P) by exchanging $\beta \leftrightarrow L_1$, we will show the
results in the (AP,P) case only.)

The order parameters of this theory are the Wilson loop operators
winding around each cycle:
\begin{align} 
&\Tr U= \frac{1}{N}  \Tr P \exp\left(  i \int_0^{\beta}A_0 dt \right), \quad
\Tr V= \frac{1}{N}  \Tr P \exp\left(  i \int_0^{L_1}A_1 dx_1 \right), \nn
&\Tr W= \frac{1}{N}  \Tr P \exp\left(  i \int_0^{L_2}A_2 dx_2 \right).
\label{Polyakov3}
\end{align} 
If the gravity solution has a contractible cycle (i.e.  it wraps
around a `cigar'), the expectation value of the corresponding Wilson
loop operator is non-zero.  
If the solution is localized on a cycle,
then also the expectation value of the corresponding Wilson loop
operator is non-zero. However, if the solution wraps around a
non-contractible cycle, the expectation value of the corresponding
Wilson loop operator vanishes \cite{Aharony:2004ig}.
 
In order to derive the phase structure corresponding to figure \ref{fig-phases}, we
will evaluate the phase structure of supergravity for each boundary
condition by changing $\beta$ and $L_1$ with a fixed $L_2$, which is
related to a cut off scale of the 2 dimensional gauge theory
(\ref{d=2}) (see footnote \ref{ftnt-strong-coupling}).

From now on, we use units such that $\lambda_3=1$.

\subsubsection{D2 on (AP,AP,AP) torus}

We consider the phase structure of $D2$ branes on a 3-torus
with (AP,AP,AP) boundary conditions. 
In this case, three solutions appear: $D2_{ \beta(L_1,L_2)},D2_{ L_1(\beta,L_2)}$ and $D2_{ L_2(\beta,L_1)}$. 
(Recall the notation at the end of section \ref{sec-Dp-sol}.)
In this case, the theory does not have the P circle and only the SS transition (\ref{ss-value}) happens.
 The phase structure is shown in figure \ref{fig-APAPAP-L}.

\begin{figure}
\begin{center}
\includegraphics[scale=.75]{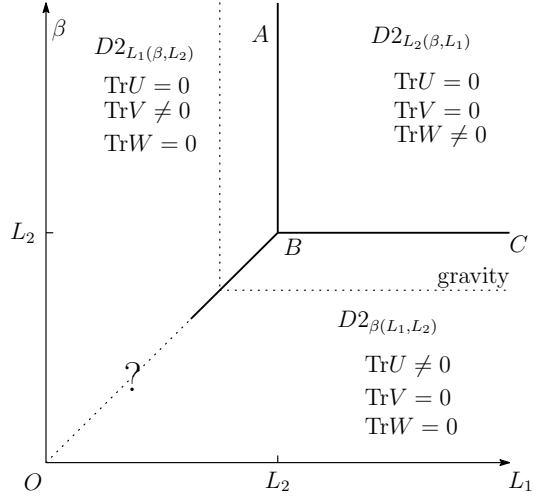}
\caption{Phase structure of the D2 brane on $T^3$ with (AP,AP,AP)
  boundary condition for large $L_2$.
  The gravity analysis is valid above the dotted line.
  }
\label{fig-APAPAP-L}
\end{center}
\end{figure}

As we have argued in appendix \ref{sec-gr-valid}, we need to check the
validity of the gravity solutions.  Let us consider the solitonic
solution $D2_{L_2(\beta,L_1)}$.  From (\ref{cond-string}), $L_2 \gg 1$
is required (in units where $\l_3=1$) to suppress the stringy
excitations around the tip of the cigar.  
We need to check the condition related to the winding modes also.  In
order to suppress the winding modes, $L_1^{3/2},~\beta^{3/2} \gg L_2$
are required from (\ref{cond-winding}).  
The phase boundary $AB$ and $BC$ are given by $L_1=L_2$ and $\beta=L_2$, and these conditions are satisfied on these boundaries in the large $L_2$ case. Thus the $D2_{L_2(\beta,L_1)}$ phase is always reliable, if $L_2$ is large.

Next we consider the black brane solution $D2_{\beta(L_1,L_2)}$. From
(\ref{cond-string}), $\beta \gg 1$ is required. 
Thus this solution is not reliable when $\beta \sim O(1)$.  
 We can see that the condition
(\ref{cond-winding}) for the winding modes is satisfied in the region
surrounded by the phase boundaries and $\beta \gg 1$.  Therefore the
$D2_{\beta(L_1,L_2)}$ solution is reliable in the region indicated in
figure \ref{fig-APAPAP-L}.  Similarly the solitonic solution
$D2_{L_1(\beta,L_2)}$ is reliable in the region indicated in
figure \ref{fig-APAPAP-L}.

Summing up these tests for the validity of the gravity analysis,
the phase structure is reliable if $L_2$ is large and $\beta$ and $L_1$ are above the dotted line in figure \ref{fig-APAPAP-L}. 
This is, of course, a problem, since we are interested in the results in the $L_2 \to 0$ limit as we mentioned in Section \ref{sec-gravity}.  
We will discuss this problem in appendix \ref{app-conjecture}.

\subsubsection{D2 on (AP,P,AP) torus}

In this case, 4 solutions appear: $D2_{ \beta(L_1,L_2)}, D2_{ L_2(\beta,L_1)}, D1_{ L_2(\beta)}$ and $D1_{ \beta(L_2)}$.
The phase structure for large $L_2$ is shown in figure \ref{fig-APPAP-L}.  
The phase boundaries are given as
\begin{align} 
AB: L_1=\left(\frac{C_1}{C_2}  \right)^2 L_2^{2/3}, \quad EC: \beta=L_2, \quad BO: \beta=\left(\frac{C_2}{C_1}  \right)^3 L_1^{3/2}.
\end{align} 
Note that $AB$ is the GL type transition (\ref{gl-s-value}), $BO$ is the GL transition (\ref{gl-value}) and $EC$ is the SS transition (\ref{ss-value}).

Through similar tests for the validity of gravity as before, we find
that the gravity analysis is valid only in the region above the dotted
line in figure \ref{fig-APPAP-L}.
(Again the gravity analysis in a small $L_2$ is invalid.)

\begin{figure}
\begin{center}
\includegraphics[scale=.75]{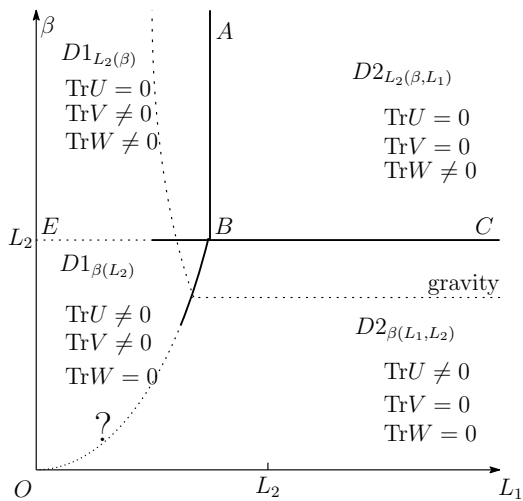}
\caption{Phase structure of the D2 brane on $T^3$ with (AP,P,AP)
  boundary condition for large $L_2$.  The gravity
  analysis is reliable only in the region above the dotted line.}
\label{fig-APPAP-L}
\end{center}
\end{figure}

\subsubsection{D2 on (P,P,AP) torus}

In this case, 4 solutions appear: $D2_{ L_2(\beta,L_1)}, D1_{ L_2(\beta)},  D1_{ L_2(L_1)}$ and $D0_{ L_2}$.
The phase structure for a large $L_2$ is shown in figure \ref{fig-phases-gr}.  
The phase boundaries are given by
\begin{align} 
&AB: L_1=\left(\frac{C_1}{C_2}  \right)^2 L_2^{2/3}, \quad BD: \beta=L_1, \quad DO: \beta=\left(\frac{C_0}{C_1}  \right)^{5/2}L_2^{1/2} L_1^{1/4} .
\end{align} 
These are GL type transitions (\ref{gl-s-value}).
Other lines can be obtained by $\beta \leftrightarrow L_1$.
The gravity analysis is valid only in the region above the dotted line
in figure \ref{fig-phases-gr} for large $L_2$.

We will adopt the phase structure in this boundary conclusion as a prediction for the gauge theory.
The reason will be explained in the next section.


\subsection{Conjectured phase diagram of 2D bosonic gauge theory}
\label{app-conjecture}

In the preceding subsections, we have obtained various phase
structures from gravity for different spin structures ({\em i.e.},
for different fermion boundary
conditions).
Since these results are valid only for large $L_2$, we need to extrapolate them to small $L_2$, where the bosonic gauge theory should appear (see footnotes \ref{ftnt-strong-coupling} and \ref{support}).

From the viewpoint of the two-dimensional bosonic gauge theory, a
holographic correspondence with the various gravity phase diagrams described
in this section is problematic since the latter have a dependence on 
fermionic boundary conditions, while no such dependence obviously
exists for the two-dimensional gauge theory  (such dependences, however, 
exist for the three-dimensional SYM theory, which is more
directly related to the gravity description). 
Indeed one such problem  was pointed out in \cite{Aharony:2006da}.
This argument is further developed in \cite{Mandal:2011ws}.
The arguments presented in these papers indicate that the phase transition in the bosonic gauge theory cannot be regarded as a continuation of the SS transition of gravity.
Thus, in order to have a smooth continuation of phase boundaries
between the holographic description and the two-dimensional gauge theory, we should avoid choosing the (AP,AP,AP) and (AP,P,AP) boundary conditions, in which the SS transition appear.
For this reason, we choose the (P,P,AP) case to read off the predicted
phase structure from gravity.

Note that all the transitions in the (P,P,AP) case are of
the GL-type, and are supposed to be first order phase transitions
\cite{Aharony:2004ig, Harmark:2007md, Hanada:2007wn}.  In this case,
in addition to the uniformly distributed solitonic $(p-1)$ brane and
localized solitonic $(p-1)$ brane discussed in \ref{app-GLtype},
non-uniformly distributed solitonic $(p-1)$ brane, which is always
unstable, appears.  These three solutions would correspond to the
uniform, non-uniform and gapped distribution in the gauge theory (see figure (\ref{fig-3-phase})).  Indeed, the free energies of these gravity
solutions are expected to satisfy a  similar relation to those
in the gauge theory 
shown in figure \ref{fig-potential} through numerical study in general
relativity \cite{Kudoh:2004hs}.  This fact also supports our prediction
from the gravity analysis.

\end{document}